\documentclass{ar-1col}
\usepackage[numbers]{natbib}
\usepackage{url}
\usepackage{amsmath}
\usepackage{lineno}
\doi{10.1146/annurev-nucl-102020-022253}

\begin{document}
%\linenumbers
% Page header
\markboth{Arrington et al.}{Short-range correlations}
\title{Progress in understanding short-range structure in nuclei: an experimental perspective}

%Authors, affiliations address.
\author{John Arrington,$^1$ Nadia Fomin,$^2$ and Axel Schmidt$^3$
\affil{$^1$Lawrence Berkeley National Laboratory, Berkeley, CA94720; email: jarrington@lbl.gov}
\affil{$^2$Department of Physics and Astronomy, University of Tennessee, Knoxville, TN, 37996; email: nfomin@utk.edu}
\affil{$^3$Department of Physics, The George Washington University, Washington, DC, 20052}}

%Abstract
\begin{abstract}
\textbf{High-energy electron scattering is a clean and precise probe for measurements of hadronic and nuclear structure, with a key role in understanding the role of high-momentum nucleons (and quarks) in nuclei. Jefferson Lab has dramatically expanded our understanding of the high-momentum nucleons generated by short-range correlations, providing sufficient insight to model much of their impact on nuclear structure in neutron stars, and in low- to medium-energy scattering observables including neutrino oscillation measurements. These short-range correlations also appear to be related to the modification of the quark distributions in nuclei,  and efforts to improve our understanding of the internal structure of these short-distance and high-momentum configurations in nuclei will provide important input on a wide range of high-energy observables.}
\end{abstract}

%Keywords, etc.
\maketitle

%Table of Contents
\tableofcontents

\maketitle

%%%%%%%%%%%%%%%%%%%%%%%%%%%%%%%%%%%%%%%%%%
%
% Suggested conventions (can discuss/change). Generally chosen to simplify notation/latex.
%
% GeV, GeV$^2$ for energy, mass. GeV/c only for momenta
% Do not use math mode for "A", n, p, N, Z, np-SRC, etc... $^2$H instead ofD for deuteron.
% 
%
%%%%%%%%%%%%%%%%%%%%%%%%%%%%%%%%%%%%%%%%%%

\section{Introduction}

The structure of atomic nuclei is well described by mean-field calculations, which can reproduce a wide range of static properties and low-energy scattering observables over the full range of the periodic table. For some observables at higher energies scales, the contribution of nucleons with momenta above the Fermi momentum, $k_F$, become more important and it has been a long-time goal of higher energy nuclear structure experiments to isolate and study these contributions. While the contribution from nucleons above $k_F$ to the momentum distribution of nucleons in nuclei is low, the fact that they are at large momenta and, as we will see below, are associated with large excitation of the residual (A-1) nucleus, make them an important piece of nuclear structure and give them a dominant role in certain high-energy scattering observables.

These high-momentum nucleons are generated by the short-distance part of the nucleon-nucleon (NN) potential - the strong tensor attractive component for separations near 1~fm, and the repulsive core at shorter distances. Hard NN interactions at short distances generate pairs of nucleons with large relative momenta, but small total momentum, known as short-range correlations (SRCs). This leads to a picture of the nucleus where nucleons with momenta below the Fermi momenta are dominated by the mean-field structure of complex many-body nuclei, while structure at larger momenta is dominated by the two-body physics associated with the short-distance NN interaction.

Because of the underlying two-body physics, these contributions are similar in scattering from heavy nuclei or scattering from the deuteron. In principle, this makes is possible to combine information from a variety of different observables in different nuclei to provide an improved understanding of the nature of SRCs, and to map out their strength and structure in light and heavy nuclei.  

Electron scattering provides a clean and powerful probe of nuclear structure, due to the combination of the point-like nature of the probe, the well understood interaction via virtual photon exchange, and the fact that the electromagnetic interaction that allows a uniform probe of the entire nuclear volume. However, several complications arise when trying to isolate the high-momentum components of the nuclear wave-function, and our understanding of the structure of SRCs comes from a combination of different measurements, each yielding unique but limited information, but together providing an ever clearer picture of these important but poorly understood component of nuclear structure. 

For the early theoretical foundations for SRCs, see Refs.~\cite{Frankfurt:1981mk, Frankfurt:1988nt}. Additional examinations of the experimental considerations for probing SRCs are presented in~\cite{Sargsian:2002wc, Arrington:2011xs}. More recent reviews include~\cite{Arrington:2011xs, Hen:2016kwk, Fomin:2017ydn}, but significant experimental progress has been made since these works were published. In this review, we will provide a brief overview of early studies, a more complete view of recent experimental progress and efforts to tie together information from different experiments. We will also highlight key remaining questions and future experimental work.

\section{Measurements of high-momentum nucleons and SRCs}\label{sec:high-p-challenges}

One important consideration is the fact the the momentum distribution of nucleons in a nucleus is not an experimental observable. Most experiments work in the framework of the Plane-Wave Impulse Approximation (PWIA), where the A(e,e'p) cross section is connected to the spectral function, which accounts for both the momentum and energy distribution of the nucleons. In this case, the spectral function (or momentum distribution) is directly related to the cross section in the PWIA assumption, so direct comparison of the cross section with calculations based on a momentum distribution of spectral function are possible. This is the approximation we make when we discuss using the cross section data to study the momentum distribution or spectral function of nuclei.

We begin by highlighting early attempts to probe the distribution of nucleons at large momenta using A(e,e'p) measurements and note some of the issues with interpreting these data. We then summarize how SRCs were isolated in inclusive scattering, with a handful of precision measurements mapping out the relative contribution of SRCs in nuclei.  The following section~\ref{sec:SRCstructure} will discuss studies of the momentum distribution of SRCs in nuclei, as well s their internal isospin and momentum structure.

\subsection{Challenges in isolating/measuring high-momentum nucleon}

In the Plane-Wave Impulse Approximations (PWIA), one can treat the A(e,e'p) single-nucleon knockout reaction as billiard-ball scattering from a single proton with (A-1) spectator nucleons. Knowing the electron beam energy and detecting the scattered electron and knocked-out proton, there is enough information to reconstruct the initial momentum (and energy) of the struck proton, referred to as the missing momentum ($p_m$) and missing energy ($E_m$). In principle, this should allow extraction of the proton distribution in the nucleus, but effects beyond the PWIA make this a significant challenge.  

The potential to have large contributions from meson-exchange currents (MEC) and final-state interactions (FSIs) make probing the momentum  distribution difficult, especially for large values of missing momentum, even for light nuclei~\cite{Sargsian:2002wc,Arrington:2011xs}.  Proton knockout measurements from the deuteron, measured at Jefferson Lab over a wide kinematic range~\cite{CLAS:2007tee, HallA:2011gjn}, have been used to examine final-state interactions, with comparisons to the Generalized Eikonal Approximation~\cite{Sargsian:2009hf} showing that this formalism provides a reasonable description of FSIs over a large kinematic range. However, recent measurements have pushed such studies to higher four-momentum transfer, $Q^2$, and larger $p_m$, focusing on regions of minimal FSIs, and observed that none of the calculations reproduce the data above $p_m=700$~MeV/c. Because of this, and the lack of high-precision systematic studies, it is not yet clear to what degree the FSI corrections can be minimized or corrected for. Because of this, SRC studies have generally moved away from A(e,e'N) measurements, with the exception of comparisons of proton and neutron knockout, where these effects are assumed to have significant cancellation. These will be discussed in more detail in Section~\ref{sec:iso-2Nknockout} where we discuss progress in exclusive studies.

Inclusive scattering has also been used to try to study the momentum distribution of nucleons.  The shape of the quasielastic (QE) is driven by the e-p and e-n elastic cross sections smeared by the nucleon momentum distributions~\cite{Benhar:2006wy}, with scattering below the kinematic threshold for stationary nucleon, allowing for clean separation of QE from inelastic scattering. Under certain assumptions, $y$ scaling~\cite{Day:1987az, Arrington:1998ps,Benhar:2006wy} provides a connection between the QE cross section and the underlying momentum distribution. However, this relies on the PWIA, and MECs and FSIs can yield large corrections at low $Q^2$. It also assumes a final spectator (A-1) nucleus in an unexcited state, which is a poor approximation for scattering from an SRC within a nucleus. Model-dependent corrections (e.g. $y_{CW}$ or $y*$ scaling~\cite{CiofidegliAtti:1999is,Arrington:2003tw}) have been proposed to account for this, but it is difficult to quantify the uncertainty in such an approach. We note, however, that in QE scattering from the deuteron at high $Q^2$, the $y$-scaling assumptions appear to be reasonable, and momentum distributions extracted from such data are in good agreement with momentum distributions calculated from realistic NN potentials, as shown in~\cite{Fomin:2011ng}.

\subsection{Experimental Signatures of SRCs}

As noted in the previous section, the shape of the QE peak is driven by the distribution of protons and neutron in the nucleus. One can decompose the cross section into contributions based on the structure (in particular high-momentum nucleons) in the breakup of the nucleus. In the naive SRC model~\cite{Frankfurt:1993sp}, the cross section is separated into scattering from single nucleons in the nucleus with the spectator in an unexcited or minimally-excited states, scattering from two-nucleon SRCs where the final state is dominated by the high-momentum spectator and an (A-2) residual nucleus, and contributions from multi-nucleon SRCs where multiple nucleons in the initial state have large momenta. The bulk of the cross section comes from the first term, which represents the shell structure of the nucleus. Contributions from SRCs are assumed to be significantly smaller, with the contributions falling as the number of high-momentum nucleons increases. Therefore, initial studies of SRCs assumed that only 2N-SRCs would have significant contributions, and made measurements focused on scattering from high-momentum nucleons ($k > k_{Fermi}$) to suppress the contributions from the shell-model structure.

\begin{figure}[htbp]
\centering
\parbox{5.8cm}{
\includegraphics[width=5.7cm, height=6.8cm, trim={1mm -2mm 0mm 0mm}, clip]{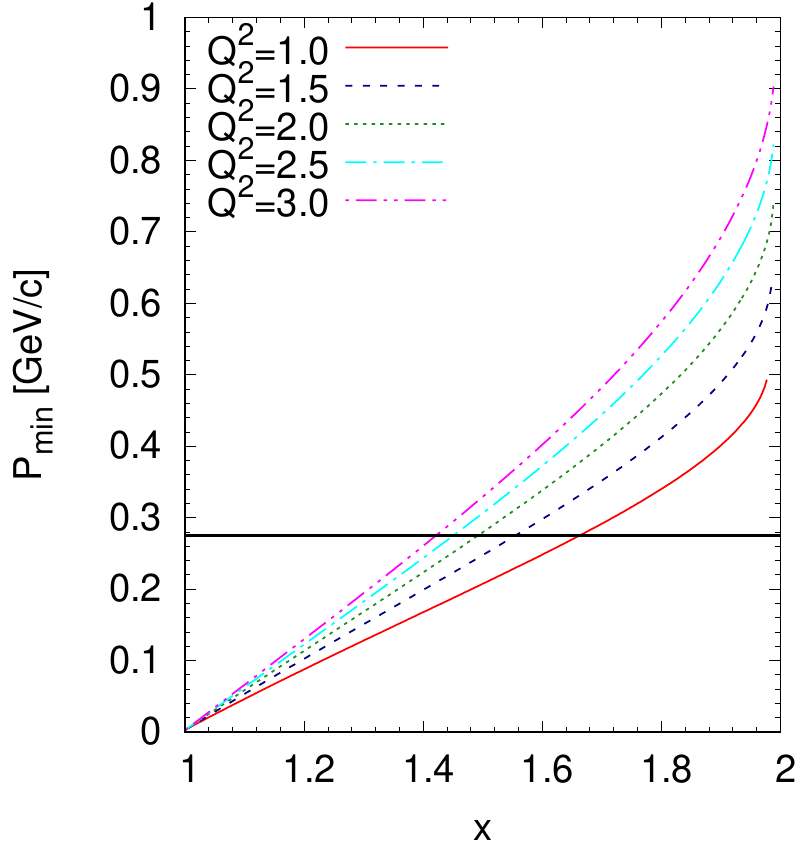}}
\parbox{6.8cm}{
\includegraphics[page=2, width=6.7cm, trim={3mm 0mm 3mm 6mm}, clip]{plots/fsds_fig3_mod.pdf}}
\caption{(Left) Minimum nucleon momentum for quasielastic scattering from a nucleon as a function of $x$ and $Q^2$ (Right) Fe/D inclusive cross section ratios (per nucleon) from SLAC over a range of $Q^2$ values; figure adapted from Ref.~\cite{Frankfurt:1993sp}.}
\label{fig:a2_thresh}
\end{figure}

Electron scattering from a stationary proton is kinematically forbidden for energy transfer, $\nu$, below that for elastic e-p scattering, corresponding to $x=Q^2/(2M\nu)=4$. In a nucleus, scattering at $x>1$ is can occur because of the motion of the nucleon in the nucleus, with larger values of $x$ corresponding to larger initial nucleon momenta. Therefore, by selecting scattering at $x>1$, one can set a minimum initial nucleon momentum that depends on $x$ and $Q^2$, as shown in the left plot of Fig.~\ref{fig:a2_thresh}. Measurements at modest-to-large $Q^2$ values ($>$1-2 GeV$^2$) will minimize final state interactions, while the low energy transfer required to reach large $x$ suppresses inelastic scattering. Thus, it was predicted that it should be possible to isolate scattering from the high-momentum nucleons in SRCs by requiring that $x$ and $Q^2$ be large enough that scattering from nucleons below the Fermi momentum is forbidden~\cite{Frankfurt:1981mk, Frankfurt:1988nt}. 

In this region, the inclusive scattering from any nucleus is driven by scattering from 2N-SRCs, generated by the same underlying two-body physics. The scattering in this region should therefore show a universal behavior, yielding a nuclear cross section ratio in the SRC-dominated region that is independent of both $x$ and $Q^2$. The first extensive set of data to examine this prediction was from SLAC~\cite{Frankfurt:1993sp}, which found a plateau in the A/D cross section ratio for $x>1.4$ and $Q^2>1.4$~GeV$^2$ for $^3$He, $^4$He, $^{12}$C, $^{27}$Al, $^{56}$Fe, and $^{197}$Au. The results for the Fe/D ratios are shown for a range of $Q^2$ values the right panel of Fig.~\ref{fig:a2_thresh}. High-momentum nucleons ($k > k_{Fermi}$) are accessible at lower values of $x$ as $Q^2$ increases, as seen in the right panel of Fig.~\ref{fig:a2_thresh}, yielding a larger plateau region in the high-$Q^2$ A/D ratios.
These data confirmed the prediction of identical behavior in light and heavy nuclei as a function of $x$ and $Q^2$ for measurements made in the SRC-dominated region, supporting the naive SRC model. Taking the cross section for scattering from SRCs in heavy nuclei to be identical to scattering from a deuteron, the quantity $a_2$ - the A/D ratio in the plateau region - was taken as an approximate measure of the relative contribution of SRCs relative to that in the deuteron.

\begin{figure}[htbp] 
\centering
\parbox{6.3cm}{
\includegraphics[width=6.2cm,height=4.2cm]{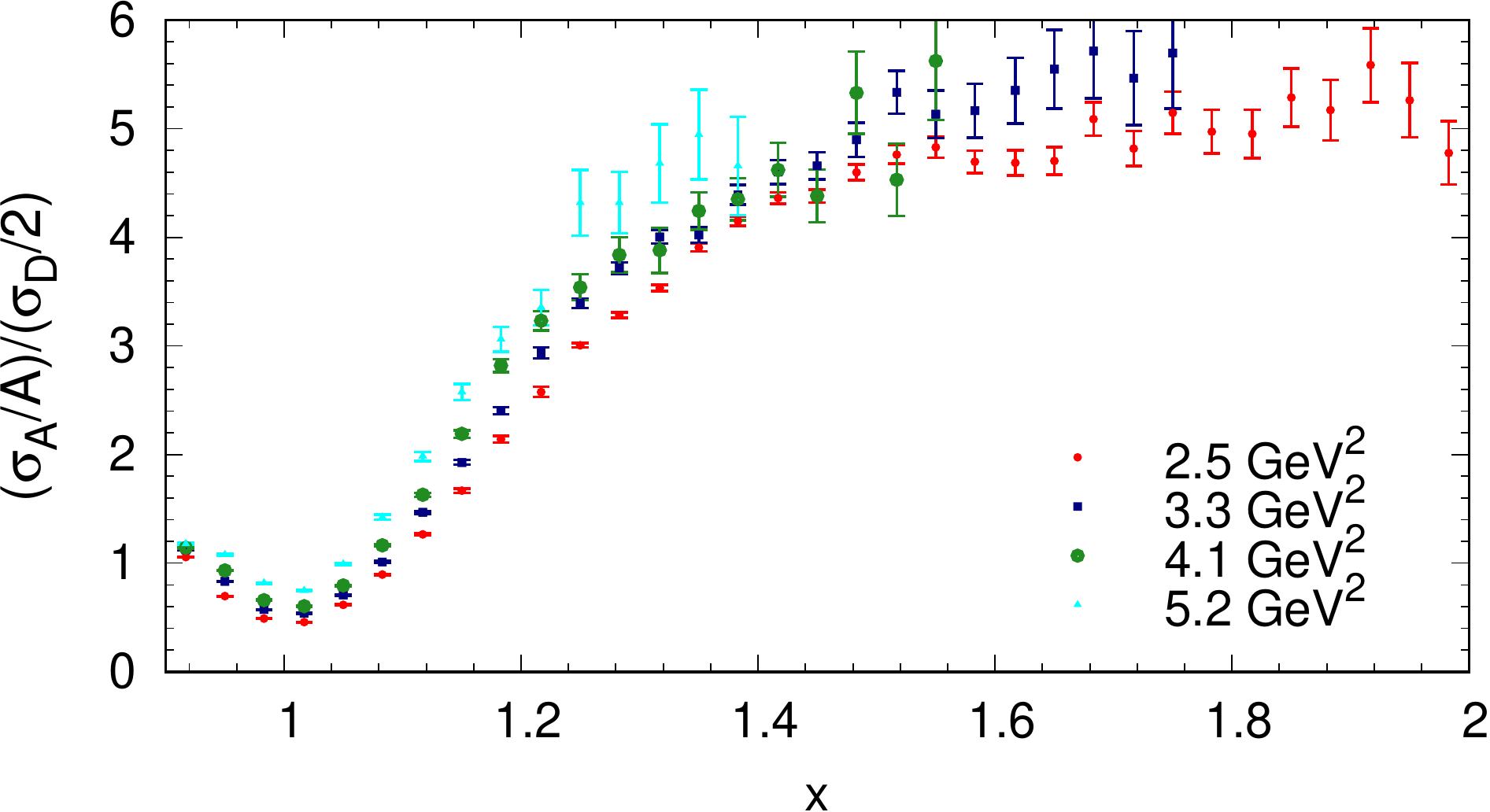}}
\parbox{6.3cm}{
\includegraphics[trim={19mm 30mm 82mm 67mm}, clip, width=6.2cm,height=4.2cm]{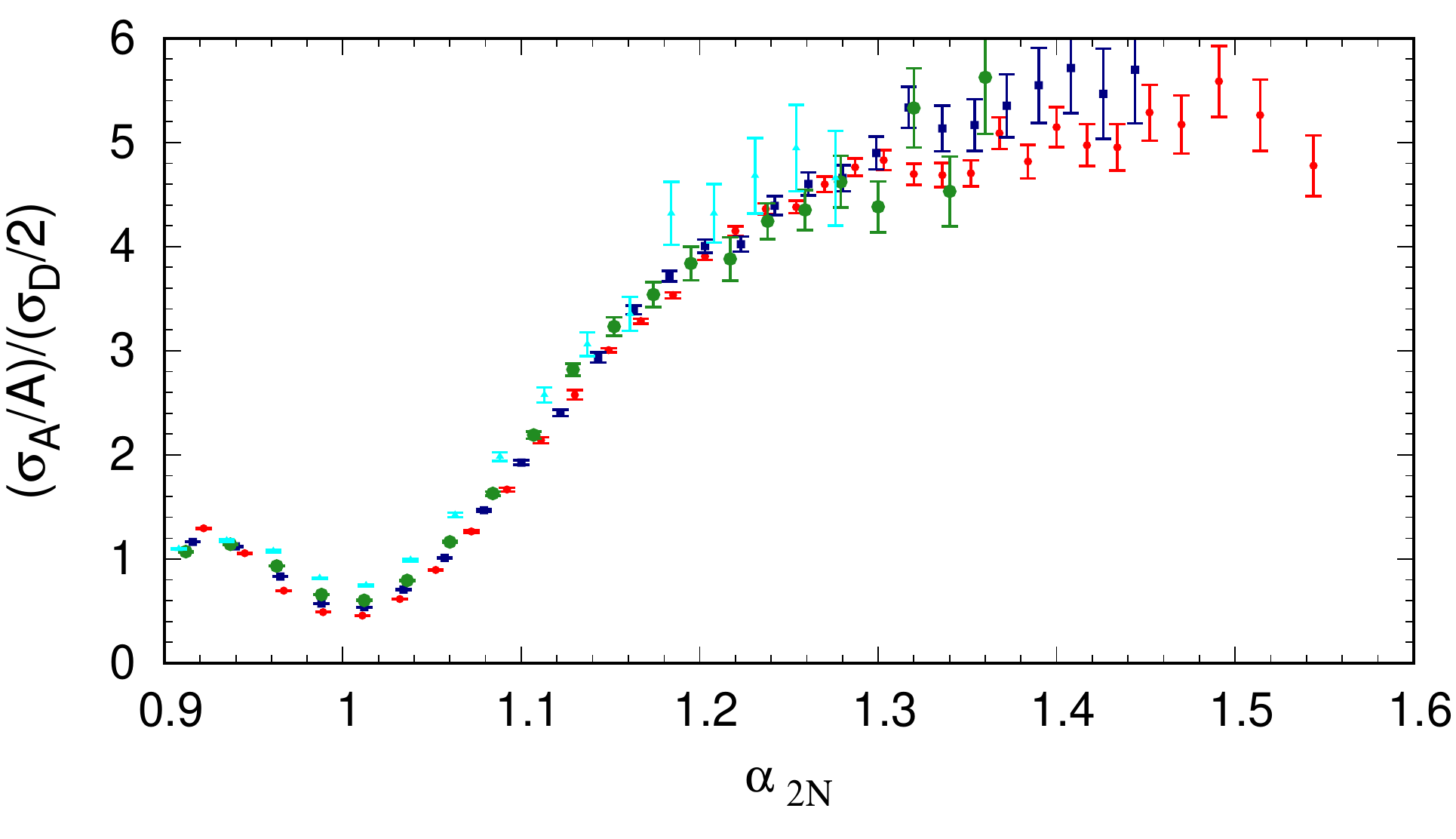}}
\caption{$^{12}$C/D per-nucleon cross section ratios from~\cite{Fomin:2011ng} plotted as a function of $x$ (left panel) and $\alpha_{2N}$ (right panel). $Q^2$ values for $x=1$ for the different data sets are shown in the left plot; the data is nearly independent of $Q^2$ as a function of $\alpha_{2N}$.}
\label{fig:x-vs-alpha}
\end{figure}

At very large $Q^2$, $x$ can be directly equated with a specific minimum nucleon momentum, while going to lower $Q^2$ yields a reduction in the range of nucleon momenta associated with the fixed value of $x$. Thus, the use of $x$ as a proxy for the initial nucleon momentum an approximation at lower $Q^2$ values, yielding a $Q^2$ dependence in the region where the scattering is not dominated by SRCs. This $Q^2$ dependence can be largely removed by using the light-cone variable $\alpha _i$, rather than $x$,  as a surrogate for nucleon momentum. The light-cone momentum fraction $\alpha _i$ is analogous, but is calculated for a bound nucleon in a nucleus, representing fraction of the total nuclear momentum, carried by the nucleon on the light-cone.  The general form is $\alpha _i=\frac{p_{i\_}}{p_{A\_}/A}$, where $p_i\_$ is the longitudinal component of the momentum on the light-cone for the bound nucleon, and $p_A\_$ is the same for the nucleus. In the case of a 2N-SRC, the expression becomes: 
\begin{equation}
    \alpha _{2N}=2-\frac{q_{min}+2M}{2M}\left( 1+\frac{\sqrt{W^2-4M^2}}{W} \right) ~~,
\label{eqn:alpha}
\end{equation}
where $q_{min}=\nu -|\vec{q}|$, $W^2=\sqrt{-q^2+4M\nu+4M^2}$, and $M$ is the proton mass. When the 2N-SRC cross-section ratios are examined as a function of $\alpha _{2N}$, rather than $x$, as shown in Fig.~\ref{fig:x-vs-alpha}, the $Q^2$ dependence largely removed, and the onset of scaling is more consistent.  Note that for $x \approx 1$, the larger values of $Q^2$ show deviations from scaling even when shown vs $\alpha _{2N}$, as the large inelastic contributions begin to become significant in the cross section. Because the larger smearing in heavy nuclei increases the contribution from the inelastic, the A/D ratio increases for $x\approx1$ and large $Q^2$, leading to the increase in the ratio.

Once scaling of the cross-sections is observed, there are a couple of lingering obstacles to address in attempting to quantify the contribution of SRCs in these nuclei. A common concern is the effect of FSIs, which, despite falling off quickly with increasing $Q^2$ can have a lingering effect. Most calculations agree that the FSIs at large $Q^2$ are limited to interactions between the nucleons in the SRC pair, and as such, cancel in the cross section ratios at $x>1$~\cite{Frankfurt:1981mk, Frankfurt:1988nt, Frankfurt:1993sp, CiofidegliAtti:1994ys, CiofidegliAtti:1995qe, Arrington:2011xs}, and this assumption is applied universally in the interpretation of the inclusive ratios in terms of SRC contributions.  High-precision $x>1$ inclusive deuteron data from Hall C (Ref.~\cite{Fomin:2011ng}) were used to extract the underlying momentum distribution at several kinematic settings based on the $y$-scaling approach which assumes no FSI.  The results across settings were consistent, showing no $Q^2$ dependence in the extracted momentum distributions, even in the SRC-dominated region. In the presence of large FSIs for $x>1.4$, the momentum distribution would be distorted, which is not observed. While this supports the idea the effect of FSIs is small, it is difficult to set a precise limit given, among other things, the uncertainty in the strength of the momentum distribution for $k > k_F$.

Having established that the inclusive ratios can isolate and quantify the presence of SRCs in nuclei, later Jefferson Lab experiments compared heavy nuclei to $^3$He in the SRC region~\cite{CLAS:2003eih, CLAS:2005ola}, and made precision measurements of A/D ratios in SRC kinematics for a range of light and heavy nuclei~\cite{Fomin:2011ng, CLAS:2019vsb}. In the naive SRC model, these measurements directly provide a relative measure of the cross section in the SRC-dominated region, and allow for a comparison of the strength of SRCs in various nuclei. However, there are corrections to the naive SRC model which have to be addressed for a quantitative interpretation of SRCs, as described in the next subsection.

\subsection{Nuclear Dependence of SRCs}\label{sec:adep}

\begin{figure}[htbp] 
\parbox{6.9cm}{
\includegraphics[width=6.8cm]{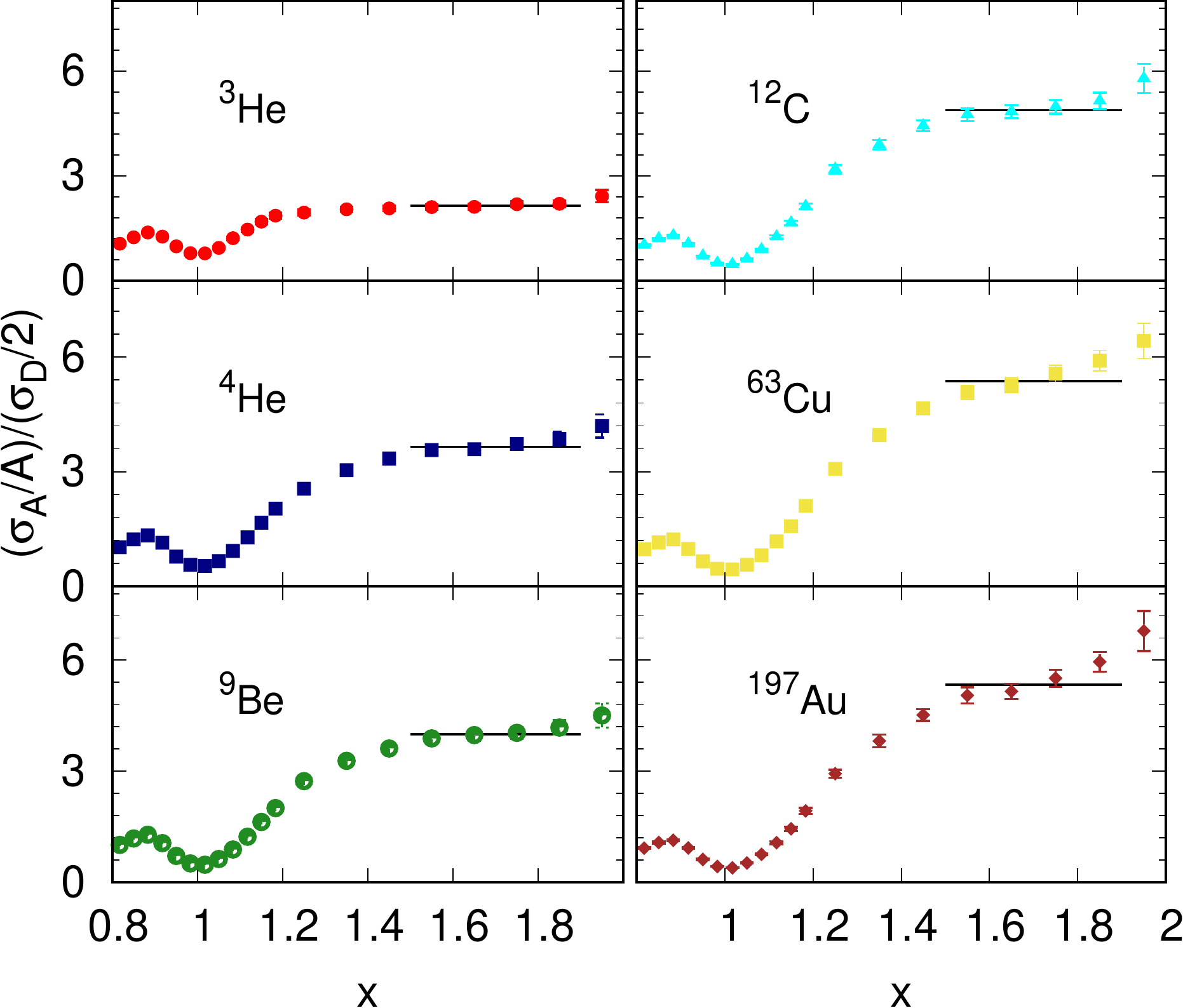}
}
\parbox{6.1cm}{
\includegraphics[width=6.0cm, trim={0mm -10mm 0mm 0mm}, clip]{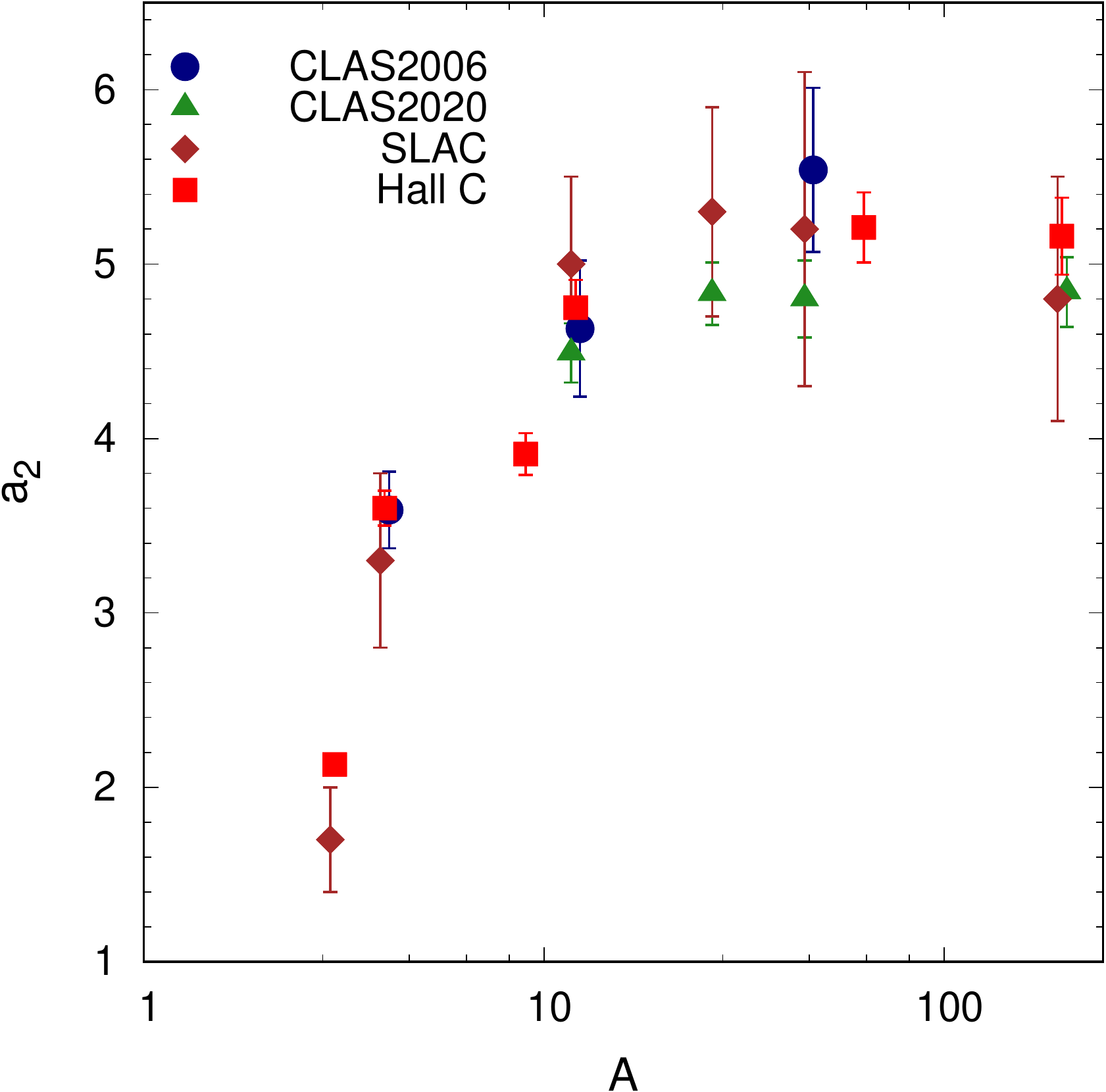}}
\caption{(Left) Cross section ratios from JLab E02-019~\cite{Fomin:2011ng} (Right) Extracted values of $a_2$ vs A from all measurements; some values offset slightly in A to make the results more visible. Note that smearing corrections have been estimates to decrease the relative number of SRCs from the extracted $a_2$ value by about 10\% for $^3$He, and 20\% for A~$\geq 12$.}
\label{fig:a2_ratios}
\end{figure}

Scaling of inclusive A/D cross-section ratios at $x>1$~\cite{Frankfurt:1993sp,Fomin:2011ng, CLAS:2019vsb} was taken to be the signature of the presence of high-momentum nucleons, born in SRCs. This ratio, taken in the scaling region (which depends on kinematics), is referred to as $a_2$.  However, $a_2$ was also extracted from A/$^3$He ratios~\cite{CLAS:2003eih, CLAS:2005ola}, combined with $^3He/D$ data and/or calculations to extract the contribution of SRCs relative to the deuteron. The left plot in Fig.~\ref{fig:a2_ratios} shows the A/D ratios for a wide range of nuclei from Ref.~\cite{Fomin:2011ng}. One can see that the scaling sets in earlier for light nuclei, while heavier nuclei require larger $x$ to isolate nucleons with momenta above the larger Fermi momentum associated with heavy nuclei.

The extracted values of $a_2$ from all inclusive measurements are shown in the right panel of Fig.~\ref{fig:a2_ratios}, with the A/$^3$He ratios converted to A/$^2$H using $^3$He/$^2$H=2.12$\pm$0.06 - the average of refs.~\cite{Frankfurt:1993sp,Fomin:2011ng}. While there are some systematic disagreements between different data sets, they do not appear to be significant given the sizes of the uncertainties, especially given that there will be a scale uncertainty on all measurements from a given data set associated with the use of common set of deuterium data (or $^3$He for~\cite{CLAS:2005ola}. Early works assumed that $a_2$ would scale with the average nuclear density (or A$^{-1/3}$ as an approximation to nuclear density), but the results of Ref.~\cite{Fomin:2011ng} showed that $^9$Be was an outlier~\cite{Arrington:2012ax} from this model, and the the details of the nuclear structure are important. For heavier, the ratio is approximately consistent, consistent with the idea of an effect that saturates in heavy nuclei with roughly 5 'nearest neighbors' able to generate SRCs with any given nucleon (implying roughly 10 'nearest neighbors', only half of which are available to form SRCs assuming np dominance). However, because the heavier nuclei have significant neutron excess, it is difficult to cleanly separate A dependence from isospin effects for medium-to-heavy mass nuclei.

Some inclusive SRC ratio measurements~\cite{CLAS:2003eih,CLAS:2005ola} assumed equal probability for all isospin SRC pairs (nn, np, pp). This meant that those data for non-isoscalar nuclei were corrected for the $a_2$ determination. While heavier nuclei also have pp- and nn-SRCs, their contributions are significantly smaller, as discussed in Sec.~\ref{sec:isospin}. Therefore, the data presented in Fig.~\ref{fig:a2_ratios} are shown without the isoscalar corrections to the cross sections that applied in some workds to try to account for pp and nn contributions~\cite{CLAS:2003eih}. It is worth noting that the momentum distribution for nucleons in np-SRCs is very different from pp-SRCs~\cite{Schiavilla:2006xx}, meaning that this becomes a more important issue when trying to make quantitative comparisons of inclusive scattering, which integrates over a range of initial nucleon momenta, and exclusive reactions which measure the ratio as a function of initial momentum. This will be discussed in more detail in Secs.~\ref{sec:iso-2Nknockout} and~\ref{sec:iso-exclusive}.

There has been significant theory work trying to model the contribution of 2N-SRCs as a function of A, and trying to estimate the relative contributions of np-SRC and pp(nn)-SRCs. A naive starting point is the total number of np pairs ($N \times Z$) or pp pairs ($Z \times (Z-1) /2$), but the ratio in the high-momentum tails does have to be proportional to the overall number of np and pp pairs, as the tensor interaction generates hard interactions between nucleons in isospin zero states~\cite{Schiavilla:2006xx, Alvioli:2007zz, Wiringa:2008dn,Wiringa:2013ala} (as discussed in detail Sec.~\ref{sec:isospin}). Thus, associating SRCs with pairs of specific spin, isospin states provides a better approximation to the data~\cite{Colle:2015ena}. These simple scaling models can be compared to values of $a_2$ extracted~\cite{Schiavilla:2006xx, Arrington:2015wja, Cruz-Torres:2019fum} based on comparing high-momentum (or short-distance) parts of realistic nuclear structure calculations~\cite{Schiavilla:2006xx, Feldmeier:2011qy, Wiringa:2013ala, Colle:2015ena, Ryckebusch:2014ann, Mosel:2016uge, Ryckebusch:2019oya}. Such approaches yield qualitatively consistent results: a significant dominance of np-SRCs or pp- and nn-SRCs, and a weak dependence of the total contribution and np/pp ratio in $A \geq 12$ Nuclei. Note that while we focus on an interpretation of SRCs and their isospin structure at the hadronic level, based on the NN interaction, a quark-level model explains np dominance by associating SRCs with diquark correlations between neighboring nucleons~\cite{West2020}, with the ratio of pp- to np-SRCs depending on the contributions of 3-quark and diquark-quark contributions in the nucleon.

A more quantitative examination of the A dependence requires going beyond the assumptions of the naive SRC model, as there are a number of effects that may make the A/D cross-section ratios ($a_2$) differ from the relative number of 2N-SRC pairs.  The general argument for inclusive measurements is that the cross-section strength at $x>1$ is due to the presence of high-momentum nucleons from short-range interactions. That means that the high-momentum tail of the momentum distribution (which is probed at $x>1$) should look similar across nuclei, with $A>2$ tails looking approximately like rescaled $A=2$ tails.  This leads to the expectation that the signature of 2N SRCs will be a plateau in the cross-section ratios. This is what is observed in the data, as Fig.~\ref{fig:a2_ratios} shows. However, there are small deviations from a perfectly flat plateau, such as a small increase as one approaches $x=2$, that are more visible in heavier nuclei. There are a couple of mechanisms at play here. One has to do with the motion of the SRC pair in the field of the other nucleons (discussed below). Another possibility is that at larger $x$ there may begin to be contributions 3N-SRCs, in the region where the 2N-SRC contributions are dropping off. The 3N-SRC contribution, relative to 2N-SRCs, is expected to grow with the size of the nucleus~\cite{Sargsian:2019joj}, leading to increasingly imperfect scaling ratios. 

A major correction comes from the \textbf{center-of-mass (CM) motion} of the 2N-SRC pair in $A>2$ nuclei. This modifies the shape of the momentum distribution and redistributes strength (from the quasielastic peak to the high-momentum tail).  The first data to be corrected for this effect were those of Ref.~\cite{Fomin:2011ng}.  The authors calculated the effect of this smearing using parametrizations of the $F(y)$ scaling function convolved with realistic parametrizations for the CM motion from~\cite{CiofidegliAtti:1995qe}. The correction was on the order of 20$\%$ for most nuclei, with a low value of 10$\%$ for $^3$He. A more recent work~\cite{Weiss:2020bkp} examines the same effect in relativistic and non-relativistic convolution models, using the Generalized Contact Formalism (GCF) to provide the structure of the SRCs to be smeared, and also examining the impact of having a larger excitation of the residual (A-2) system. However, Ref.~\cite{Weiss:2020bkp} does not provide direct comparisons of calculations with and without SRC smearing or (A-2) binding, making it difficult to determine the size of the individual effects.

For the rest of this work, we will typically refer to $a_2$ as a measure of the relative contribution of SRCs, consistent with most of the literature, although efforts are ongoing to better quantify the corrections between the cross section ratios and the relative SRC contributions.

\section{Internal structure of SRCs}\label{sec:SRCstructure}

In addition to mapping out the size of the SRC contributions in nuclei, a variety of experimental studies have focused on obtaining a more detailed understanding of the structure of SRCs in nuclei. This addresses some of the questions raised by the initial inclusive measurements: what is the isospin structure of SRCs, what is the momentum distribution of SRCs in nuclei, and what is the distribution of relative momenta within SRCs? Finally, as SRCs involve large energy scales arising from finite objects interaction strongly at short distances, is the internal structure of nucleons modified by being in these dense, energetic configurations?

While the inclusive measurements were able to confirm SRC contributions by identifying the universal behavior in all nuclei in the $x, Q^2$ range predicted by the SRC model, the earlier measurements were sensitive mainly to the overall contribution from SRCs.  They did not provide information on the isospin structure of the SRCs - the relative contribution of pp-, np-, and nn-SRCs - and did not measure the momentum of SRCs in nuclei. In both cases, corrections had to be applied based on assumptions or measurements of the isospin structure and momentum distribution of SRCs. Below, we discuss our understanding of these effects and their experimental and theoretical support.

\subsection{Measurements of center-of-mass motion of SRCs in nuclei}\label{sec:SRC-CMmotion}

Theoretical predictions and experimental results suggest that typical momenta of the center-of-mass (CM) of SRC pairs is small, i.e., on the order of mean-field momenta~\cite{Wiringa:2013ala}. Nevertheless, determining the center-of-mass motion distribution of pairs has important implications. As noted in section~\ref{sec:adep}, the inclusive measurements have to apply a correction for the motion of the SRCs inside of the nucleus to extract the relative SRC contributions in different nuclei. Furthermore, the CM momentum distribution may shed some light on how short-range correlations form. So far, experiments have aimed to quantify the width of the CM momentum distribution, and have assumed a Gaussian shape, which is broadly consistent with the data given the available statistics and resolution. For what follows, we use the following convention: for two nucleons in an SRC pair with momenta $\vec{p}_1$ and $\vec{p}_2$, the center-of-mass momentum is defined by $\vec{p}_{CM}\equiv\vec{p}_1+\vec{p}_2$, while the relative momentum is $\vec{p}_{rel.}\equiv \frac{1}{2}(\vec{p}_1-\vec{p}_2)$.

Experimental determinations have been made through two-nucleon knockout reactions. One of the earliest such measurements was performed by colliding a 6--9~GeV proton beam from a fixed carbon target in the Brookhaven EVA spectrometer and identifying SRC break-up events from the $^{12}$C(p,ppn) reaction~\cite{Tang:2002ww}. Guided by theoretical indications that the transverse directions would be far more sensitive to FSIs, Tang et al., measured the longitudinal component of the CM distribution to have a Gaussian width of $143\pm 17$~MeV$/c$, i.e., somewhat smaller than the Fermi momentum. Electron scattering experiments using the spectrometers in Jefferson Lab's Hall A measured a width of $136 \pm 20$~MeV$/c$ in the (e,e'pp) channel for carbon~\cite{JeffersonLabHallA:2007lly} and a smaller width for helium-4 in the (e,e'pn) channel of $100\pm 20$~MeV$/c$~\cite{LabHallA:2014wqo}. In both cases, the small acceptances of the spectrometers required the CM width be inferred from the measured distribution using a simulation in which the electron scatters from a nucleon in a moving pair. The result was obtained by matching the data and simulated distributions. 

\begin{figure}[htbp]
\centering
\parbox{4.8cm}{
\includegraphics[width=4.2cm, height=6.0cm] {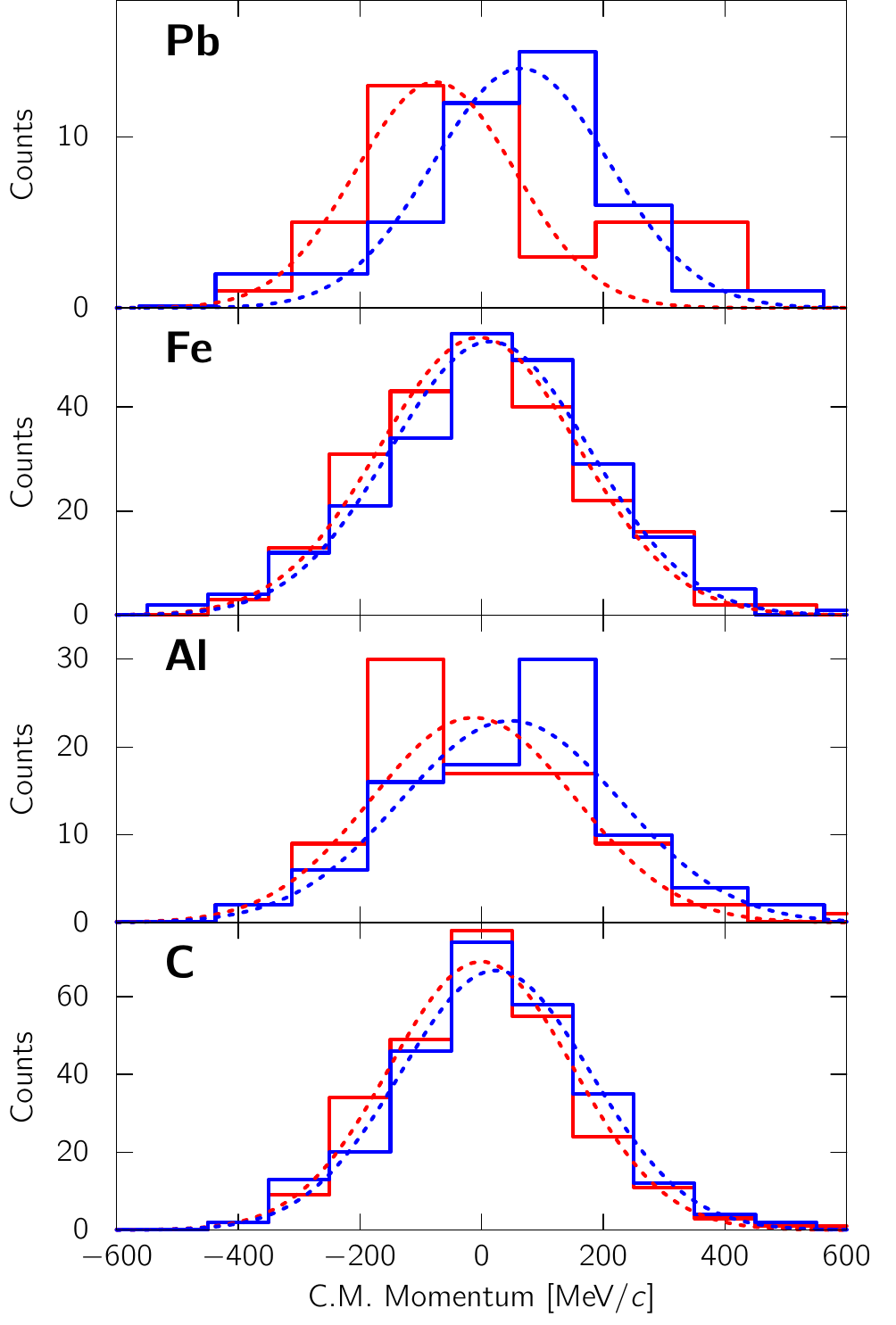}}
\parbox{7.8cm}{
\includegraphics[width=7.7cm] {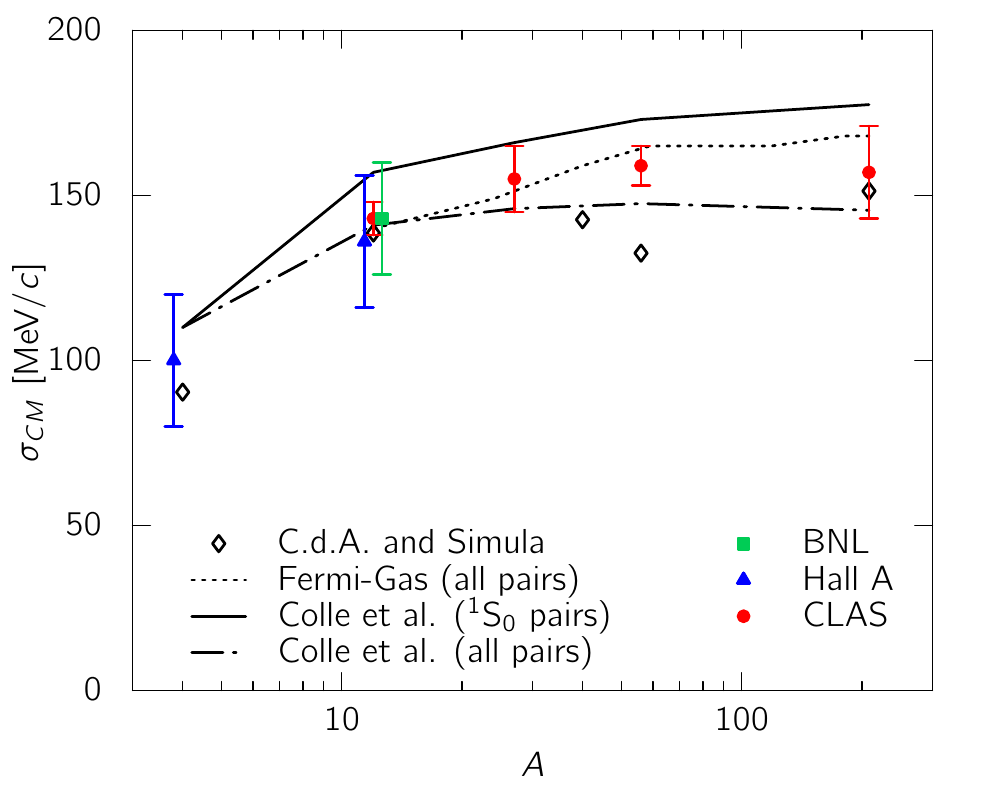}}
\caption{(Left) pp-SRC CM momentum distributions from Ref.~\cite{CLAS:2018qpc} in the direction transverse to $p_m$ in the x (red) and y (blue) directions, before correcting for the CLAS acceptance. (Right) Extraction of the gaussian width associated with SRC motion in nuclei from ref.~\cite{CLAS:2018qpc} compared to calculations from Refs.~\cite{CiofidegliAtti:1995qe,Colle:2015ena,Moniz:1971mt}; figure adapted from Ref.~\cite{CLAS:2018qpc}.}
\label{fig:2N}
\end{figure}

Measurements on a number of different nuclei have been made at CLAS, where the large acceptance is advantageous for reconstructing a multi-nucleon final state. Refs.~\cite{CLAS:2003jrg,CLAS:2010yvl} reported measurements of $^{3}$He(e,pp)n and used this to study both the relative and CM momentum of the nucleons in SRCs.
Ref.~\cite{CLAS:2018qpc} made similar measurements for additional nuclei, finding that pp-SRCs could be well described as having Gaussian CM momentum distributions with widths ranging between $143\pm 7$ MeV/c for $^{12}$C to $157\pm 14$ MeV/c for $^{208}$Pb. These results were generally consistent with the widths assumed based on mean-field momenta of the nuclei. A summary of these results is shown in the left panel of Fig.~\ref{fig:2N}.

\subsection{Isospin structure of SRCs}\label{sec:isospin}

As we discuss in the following sections, a series of measurements have made it clear that 2N-SRCs are dominated by np-SRCs, meaning that the isospin structure of SRCs is nearly identical for all nuclei, and the scattering from an SRC is roughly proportional to the sum of the e-p and e-n cross sections.

While early inclusive measurements neglected the isospin structure of SRCs, the comparison of heavy nuclei to the deuteron will be impacted by the presence of pp- and nn-SRCs, not possible in the deuteron. As noted in Sec.~\ref{sec:adep}, some measurements applied corrections to the cross section per nucleon to correct for the difference between the elastic e-p and e-n cross sections.

Multiple approaches have been used to study the isospin structure of SRCs. As noted in sec.~\ref{sec:high-p-challenges}, A(e,e'p) measurements are challenging to interpret at high missing momenta. The first measurements examining the isospin structure used detection of both the struck nucleon from the SRC and it's spectator partner. While this approach has to worry about large FSI corrections, certain contributions will cancel in the comparison of np and pp final states. Additional studies using A(e,e'p)/A(e,e'n) at large and small missing momenta and recent inclusive studies making use of the isospin structure of the target support the picture of np dominance, with each having it's own advantages and limitations.

\subsubsection{Two-nucleon knockout measurements}\label{sec:iso-2Nknockout}

Information about the isospin structure of short-range correlations can be directly accessed in measurements in which both correlated nucleons are detected. Such measurements face a couple of experimental challenges. First, since the detection of neutrons is accomplished differently than that of protons, care must be taken to control the relative acceptance and efficiency between the two particles. Second, as with all measurements with a nucleon in the final state, final-state interactions are important. For these measurements, the results are particularly sensitive to charge-exchange reactions (where an outgoing proton knocks out a spectator neutron or vice versa) or scattering from a low-momentum nucleon which rescatters from a spectator nucleon, generating a final state that's identical to scattering from a pre-existing SRC. Different experiments have used a variety of strategies to negotiate these challenges, and while these issues impact the quantitative interpretation, the body of experimental data points a clear picture of dominance of np-SRCs in the momentum range of $\approx$ 300--600~MeV/c. The evolution of isospin structure at higher momenta is an area of active research. 

\begin{figure}[htbp]
\centering
\parbox{6.3cm}{
\includegraphics[width=6.2cm, height=5.1cm, trim={0mm 0mm 0mm 2mm}, clip] {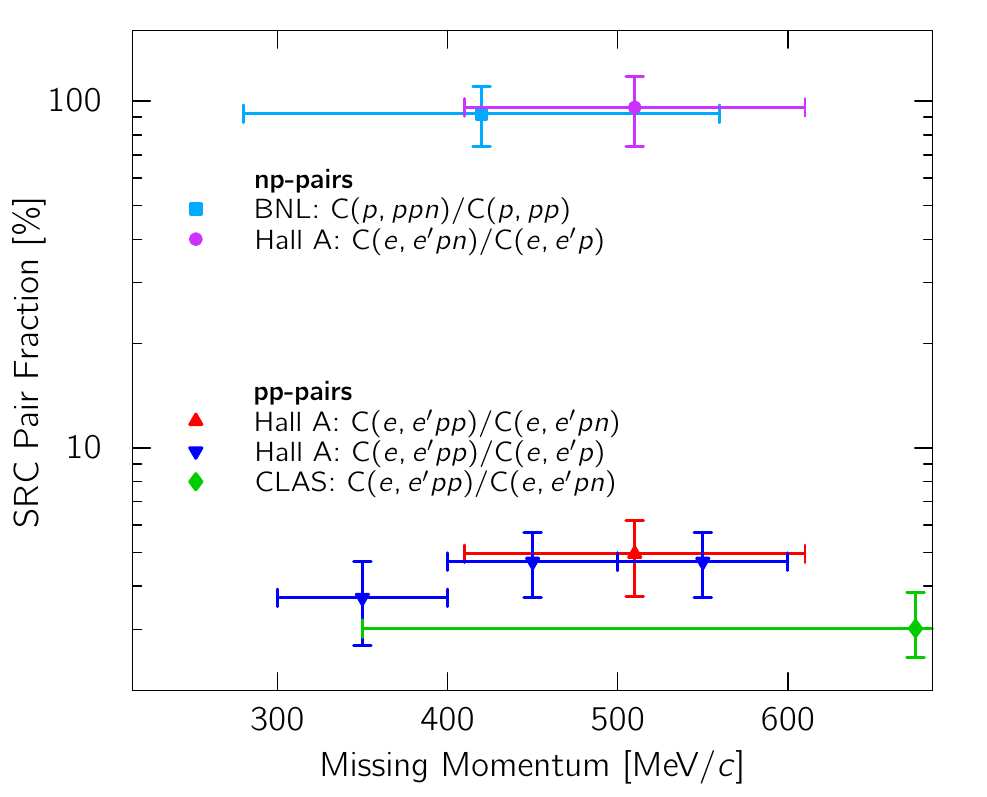}}
\parbox{6.3cm}{
\includegraphics[width=6.2cm, height=5.0cm, trim={8mm 5mm 15mm 21mm}, clip] {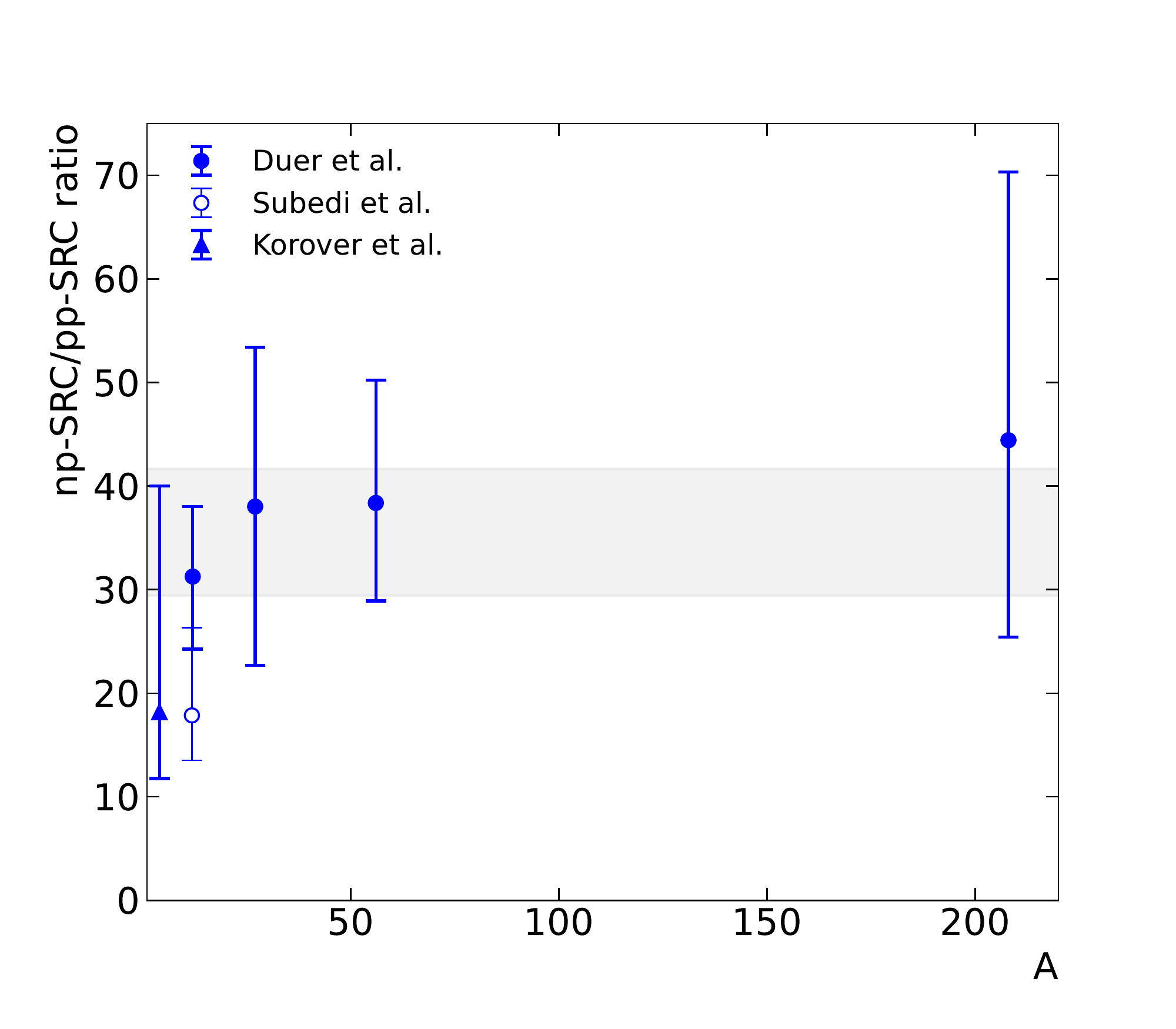}}
\caption{(Left) The pp-SRC and np-SRC fractions as a function of missing momentum from Refs.~\cite{Piasetzky:2006ai,Subedi:2008zz,JeffersonLabHallA:2007lly,CLAS:2018xvc}; figure adapted from Ref.~\cite{Subedi:2008zz}. (Right) np-SRC/pp-SRC ratio from direct measurements using two-nucleon knockout. The blue band shows the one-sigma range from the the average of all measurements excluding Ref.~\cite{Subedi:2008zz} (hollow circle) which significantly underestimated~\cite{CLAS:2018xvc} the FSI corrections.}
%John is an author and has provided permission (even though we plan to make new figures for the final publication)
\label{fig:2N-isospin}
\end{figure}

The first estimate of the isospin structure of SRC pairs was made using $^{12}$C(p,ppn) and $^{12}$C(p,pp) data collected by the EVA Spectrometer at Brookhaven, finding that correlated protons were accompanied by the emission of back-to-back neutron $92_{-18}^{+8}\%$ of the time~\cite{Piasetzky:2006ai}. This was confirmed by an electron-scattering experiment in Jefferson Lab Hall A, in which high resolution spectrometers were used to detect the scattered electron as well as a knocked-out proton in high $p_\text{miss}$ kinematics, while a third spectrometer was used to look for recoiling protons and neutrons. The pp pair fraction was inferred from both the $^{12}$C(e,e'pp)/$^{12}$C(e,e'p)~\cite{JeffersonLabHallA:2007lly} and 
$^{12}$C(e,e'pp)/$^{12}$C(e,e'pn) ratios~\cite{Subedi:2008zz}. The results of these measurements are shown in the left panel of Fig.~\ref{fig:2N-isospin}. A follow-up measurement confirmed np dominance for the $^4$He nucleus~\cite{LabHallA:2014wqo}.

An analysis from CLAS considered np-dominance in non-isoscalar nuclei~\cite{Hen:2014nza}. Since CLAS had limited capabilities for neutron detection, inferences were drawn from a measurement of the A(e,e'pp)/A(e,e'p) ratio. To compensate for the number of undetected recoil protons, double ratios were formed relative to carbon. The results showed that pp pairing is as prevalent in aluminum, iron, and lead as it is in carbon, i.e., no significant decrease increasing neutron excess. This suggests that in neutron-rich nuclei, protons have larger average kinetic energy. A follow up analysis that in which $A(e,e'np)$ events were identified via neutron hits in the CLAS electromagnetic calorimeter had consistent findings~\cite{CLAS:2018xvc}. Note that the FSI corrections applied in this work were substantially larger than applied in the initial measurement at JLab~\cite{JeffersonLabHallA:2007lly}, yielding a result $\sim$70\% larger than the initial work. The right panel of Fig.~\ref{fig:2N-isospin} shows the ratio of np-SRCs to pp-SRCs extracted from all direct two-nucleon knockout measurements.

The measurements described above have looked at scattering from nucleons in correlated pairs, with the possible additional detection of a spectator. By contrast, an analysis of $^{3}$He(e,e'pp)n data at CLAS was performed in kinematics where the spectator pair is correlated, while the struck nucleon is not~\cite{CLAS:2010yvl}. Events with pp and np pairs were identified by relative energy transfer among the detected protons and undetected neutrons. This analysis confirmed the relative scarcity of pp pairs while also observing that the effect diminishes for large pair CM momentum~\cite{CLAS:2010yvl}. It is not clear whether this is a result of the details of the $A=3$ wave function, caused by final-state effects, or produced by some other phenomenon.

\subsubsection{Single nucleon knockout}\label{sec:iso-exclusive}

One can also attempt to study the isospin structure of SRCs using single nucleon knockout. One technique is to compare relative rates of proton and neutron knockout in high $p_\text{miss}$ kinematics. A CLAS analysis of (e,e'p) and (e,e'n) in both correlated and in mean-field kinematics was undertaken to study the asymmetry dependence of pairing~\cite{CLAS:2018yvt}. It was found that while the (e,e'n)/(e,e'p) ratio in mean-field kinematics scales roughly with N/Z (after correcting for the difference between the elastic e-p and e-n cross sections), this ratio is approximately unity in SRC kinematics, a clear sign of np dominance persisting in asymmetric nuclei. The relative abundance of nucleons in correlated and mean-field states, i.e., (e,e'N)$_{SRC}$/(e,e'N)$_{MF}$, was also studied for both protons and neutrons. Since the relative acceptance for the two kinematics was very different, a double-ratio relative to carbon was used. It was found that the fraction of correlated protons steadily increases with N/Z, while the double ratio remains flat for neutrons. In a simple np-dominance picture, the neutron ratio would have a slight decrease. 

Isospin structure can also be tested in isospin mirror nuclei, i.e., where the behavior of protons in one nucleus can provide information about the neutrons in the other. An (e,e'p) experiment was conducted in Hall A on $^3$H and $^3$He pair with a goal of mapping the relative momentum distribution for protons and neutrons in the $A=3$ system. The measurement was conducted in anti-parallel, $x>1$ kinematics, for $p_\text{miss}$ in the range of 50 to 500 MeV$/c$, i.e., across the transition from the conventional mean-field to the correlated regimes. The ratio $^3$He(e,e'p)/$^3$H(e,e'p)~\cite{JeffersonLabHallATritium:2019xlj} was measured to be above 2 at low missing momentum, falling to approximately 1.4 at $p_\text{miss}=250$~MeV$/c$, consistent with both the onset of np dominance, and with PWIA calculations based on three-body spectral functions~\cite{JeffersonLabHallA:2020wrr}. However, for even larger missing momentum, the ratio rose again, inconsistent with PWIA predictions. An analysis of the absolute cross sections showed that while the $x>1$ kinematics go a long way to reducing FSIs, the can still affect the measured cross section ratio, particularly through charge-exchange reactions, which can increase the helium (e,e'p) cross section and decrease the tritium (e,e'p) cross section with a non-trivial $p_\text{miss}$-dependence~\cite{JeffersonLabHallATritium:2020mha}. While the general result is in agreement with the np-dominance picture, in general, detailed final-state interaction calculations are necessary for drawing firm quantitative conclusions about the nuclear ground state. 

\subsubsection{Inclusive studies on non-isoscalar nuclei}

Because inclusive scattering is sensitive to both proton and neutron knockout, the cross section is proportional to a combination on np-, pp-, and nn-SRCs. In general, this means that there is little sensitivity to the isospin structure of the SRCs, especially in scattering from isoscalar nuclei, and deviations from simple models of the A dependence associated with neutron excess do not provide sufficient sensitivity to look for isospin dependence. This is in part because the effects are not large, but also because it is difficult to separate the A dependence from the N/Z dependence given that the deviation of N/Z tends to grow with mass in nuclei for which SRC measurements exist. However, by comparing scattering from targets with similar mass but different isospin structure one can attempt to isolate the isospin structure of the SRCs.

The first such experiment compared $^{48}$Ca and $^{40}$Ca scattering in 2N-SRC kinematics~\cite{JeffersonLabHallA:2020wrr}. These nuclei have similar mass and should therefore equal contribution from NN-SRCs, but $^{48}$Ca should have an excess of nn-SRCs relative to pp-SRCs due to the additional neutrons. Because the e-n cross section is lower than the e-p cross section, this would yield a decrease in the cross section per nucleon in 2N-SRC kinematics, unless np- and pp-SRCs have negligible contributions. Taking into account the small difference in the expected SRC contribution for A=40 and A=48 and the ratio of the e-p to e-n cross section, the $^{48}$Ca/$^{40}$Ca per-nucleon cross section ratio was predicted to yield 0.975 in the case of 100\% np-SRCs and 0.930 for isospin-independent SRC formation~\cite{JeffersonLabHallA:2020wrr}. The measurement observed a ratio of 0.971(12), corresponding to near-total np dominance, with a 2$\sigma$ (1$\sigma$) lower limit on the np/pp enhancement factor of 2.9 (10.6). This confirmed the observation from 2N-knockout measurements while avoiding the need for large FSI corrections. However, given the very small difference between the predictions for isospin independence and np dominance, potential corrections to the assumptions used to interpret the data could impact the quantitative interpretation.

More recently, a comparison of scattering from mirror nuclei $^3$H and $^3$He was performed~\cite{tritium-src}, which has several advantages over the measurement using calcium isotopes. First, because $^3$H and $^3$He structure are nearly identical, there is next to no correction for the difference in nuclear structure apart from the isospin structure.  Second, because the fractional change in N/Z is much larger between the two nuclei, there is a much larger difference for the prediction for the $^3$H/$^3$He cross section ratio in the SRC regime: 0.74 for isospin independent vs. 1 for np dominance. This suggests that the measurement will be almost an order of magnitude more sensitive to the relative contribution of pp-SRCs.

\begin{figure}[htbp]
\centering

\parbox{6.3cm}{
\includegraphics[width=6.2cm] {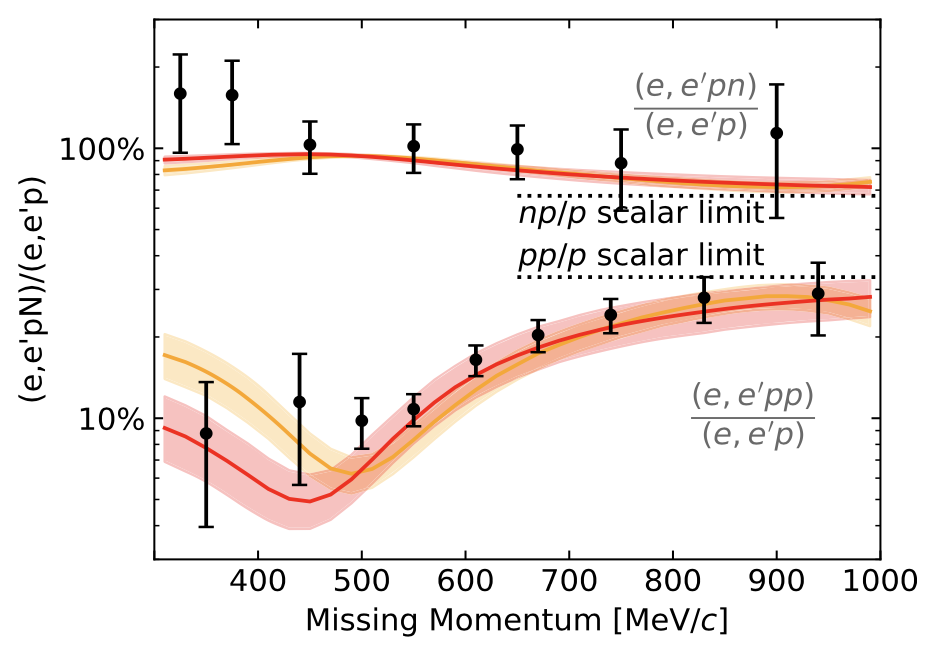}}
\caption{(Left) Measured pn/p and pp/p fractions compared to GCF calculation; Figure reproduced from Ref.~\cite{CLAS:2020rue}.}
%Axel is an author and has provided permission (even though we plan to make new figures for the final publication)
\label{fig:2N-isospin2}
\end{figure}

\subsection{SRC internal momentum distributions}\label{sec:q-relative}

As discussed in the previous sections, the ratio of np/pp SRCs varies with $p_m$~\cite{LabHallA:2014wqo}, suggesting a difference in the relative internal momentum distribution for np and pp SRCs, which is also seen in \textit{ab initio} calculations of the np and pp pair distributions in nuclei~\cite{Schiavilla:2006xx, Alvioli:2007zz, Wiringa:2008dn, Wiringa:2013ala}. As such, a more detailed understanding of the role of SRCs has to include study of the internal structure of the SRCs, which generate the observed $p_m$ dependence.  In addition, while many experiments talk about the ratio of SRCs (between nuclei, or the ratio of np to pp pairs), the measurements are sensitive to the cross section contribution of those pairs over the $p_m$ range of the measurement. As such, one must go beyond treating each measurements of the relative number of pairs, especially when comparing data from different reactions or covering different $p_m$ ranges.

In addition, measurements that include detection of one or more nucleons in the final state need to account for final-state interactions. Even in a mean field model where there are no SRCs in the initial state, simple rescattering can generate a measurement with a leading struck nucleon and a second high-momentum nucleon (above the Fermi momentum), which are indistinguishable from scattering from an SRC pair in the initial state. In addition, charge-exchange rescattering can be an exceptionally large correction, in particular when trying to isolate the very small pp-SRC contributions from the much larger np-SRC contributions.

\subsubsection{Generalized Contact Formalism}~\label{sec:GCF}

One of the challenges of interpreting the results of two-nucleon knockout measurements has been, until recently, the lack of theoretical cross section calculations with which to compare results. While there are methods for calculating the break-up of deuterium and $A=3$ nuclei, and single-nucleon spectral functions for heavier nuclei, calculations of the full two-nucleon knockout decay function for nucleons in short-range correlated pairs have proven difficult. For that reason, several recent analyses have turned to Generalized Contact Formalism (GCF), a factorized model of short-range correlations within a nucleus, to calculate fully differential two-nucleon knockout cross sections to compare to data. A theoretical derivation of GCF can be found in Ref.~\cite{Weiss:2015mba}, but the core concept is the assumption that at asymptotically short-distance or high-momentum scales, the nuclear wave function, $\Psi$, can be factorized,
\begin{equation}
    \Psi_{r_{ij}\rightarrow 0} \longrightarrow \sum_\alpha \varphi_\alpha(\mathbf{r}_{ij})A_{ij}^\alpha(\mathbf{R}_{ij},\{\mathbf{r}_{k\neq ij}\}),
\end{equation}
where $\mathbf{r}_{ij}$ is the separation vector between nucleons $i$ and $j$, $\mathbf{R}_{ij}$ is their center of mass coordinate, $\alpha$ represents the set of possible quantum numbers of a pair of nucleons, $\varphi$ represents a universal two-body wave-function that is independent of nucleus, and $A_{ij}^\alpha$ represents the wave function of the remaining $A-2$ nucleons. A similar formulation can be written in momentum space. Over these asymptotic scales, $A_{ij}^\alpha$ is approximately constant, and one can define the contact matrix,
\begin{equation}
    C^{\alpha,\beta} = 16 \pi^2 N_{ij} \langle A_{ij}^\alpha | A_{ij}^\beta \rangle,
\end{equation}
where $N_{ij}$ is the number of possible pair combinations. There are a number of selection rules that determine which off-diagonal elements can be non-zero, and in practice one only considers the diagonal terms, referred to as the ``contacts,'' $C^\alpha$, which can be interpreted as the abundance of short-range correlated pairs with quantum numbers $\alpha$, in a nucleus. 

The asymptotic behavior of the nucleons within a correlated pair is determined by the universal functions $\varphi_\alpha(\mathbf{r}_{ij})$ in position space, or $\tilde{\varphi}_\alpha(\mathbf{k}_{ij})$ in momentum space. These functions can be estimated from a given model NN potential by solving for the zero-energy solution to the 2-body Schr\"{o}dinger Equation~\cite{Weiss:2016obx}.
While different model NN potentials lead to different universal functions, comparisons with ab initio variational Monte Carlo (VMC) calculations confirm the asymptotic behavior predicted by GCF as well as the universality of pair behavior across different nuclei~\cite{Cruz-Torres:2019fum}. Comparison with VMC calculations is one of the ways that values of the contacts, i.e., the SRC pair abundances, can be estimated~\cite{Weiss:2016obx,Cruz-Torres:2019fum}.
 
In addition to modelling properties of pairs within nuclei, GCF can also be used to predict the fully differential cross section for the knockout of SRC nucleons. This is especially powerful for two-nucleon knockout, where the differential cross section has three extra differential dimensions. In single nucleon knockout, the cross section within the PWIA can be factorized
\begin{equation}
d^6 \sigma \sim K \sigma_{eN} S(E_m,\mathbf{P}_m),
\end{equation}
where $K$ is the kinematic factor, $\sigma_{eN}$ is the cross section for the electron to scatter elastically from an off-shell nucleon, and $S(E_m,\mathbf{p}_m)$ is the spectral function, describing the probability to find a nucleon within a nucleus with momentum equal to the missing momentum, $\mathbf{p}_m$, and separation energy equal to the missing energy, $E_m$. In two-nucleon knockout, the spectral function is supplanted by the two-nucleon decay function~\cite{Sargsian:2005ru}:
\begin{equation}
    d^8 \sigma \sim K' \sigma_{eN} D(E_m,\mathbf{p}_m,\mathbf{p}_r),
\end{equation}
where $K'$ is a different kinematic factor, and $D(E_m,\mathbf{p}_m,\mathbf{p}_r)$ additionally describes the probability for a second recoil nucleon to be ejected with momentum $\mathbf{p}_r$. In GCF, the decay function can be written in a factorized form:
\begin{equation}
    D(E_m, \mathbf{p}_m, \mathbf{p}_r)= \sum \limits_{\alpha} C^\alpha \left|\tilde{\varphi}^\alpha\left(\frac{1}{2}( \mathbf{p}_m- \mathbf{p}_r)\right)\right|^2
    P^\alpha_{CM}(\mathbf{p}_m+ \mathbf{p}_r)
    \delta(m_A - E_1 - E_r - E_{A-2}),
    \label{eq:gcf_decay}
\end{equation}
where $E_1 \equiv E_N-\omega$. In the above equation, $|\tilde{\varphi}^\alpha|^2$ represents a universal, i.e., nucleus-independent, two-body momentum distribution for pairs with quantum numbers $\alpha$. $P_{CM}^\alpha$ represents the center of mass momentum distribution, typically modelled by a Gaussian. A similar equation can be formulated in light-front coordinates~\cite{Pybus:2020itv}.

The tunable parameters of Eq.~\ref{eq:gcf_decay} are the contact terms, $C_\alpha$, parameters defining the CM momentum distribution (typically a single Gaussian width, $\sigma_{CM}$), and finally, any parameters describing the excitation energy of the $A-2$ system (typically denoted by a single average value, $E^*$). These parameters can either be estimated from the results of ab initio variational Monte Carlo calculations (e.g., \cite{Cruz-Torres:2019fum}), or from data. 

The usefulness of GCF for interpreting experiments is wide-ranging, and and many new avenues are under active development. First and foremost, GCF specifies a fully differential cross section for scattering from correlated nucleons. The GCF cross sections can reproduce, in $x>1$ kinematics, the multi-dimensional kinematic distributions of one- and two-nucleon knock-out~\cite{CLAS:2020mom}. Such theoretical calculations have been lacking, or used simpler approximations, in previous SRC studies. Another benefit is that GCF provides a very simple connection to underlying nuclear properties within the model, i.e., pair abundances, center-of-mass motion, etc. GCF can be used to study the sensitivity measurements of those properties, whether through the calculation of various correction factors~\cite{Pybus:2020itv}, or by performing inference from data directly~\cite{Weiss:2020bkp}. Since GCF specifies a plane-wave cross section, it can be used to identify the effects of final state interactions through significant deviations from data. Lastly, since the model is factorized, it is applicable to reactions other than electron scattering, for instance, proton scattering in inverse kinematics~\cite{BMN:2021yag}, or photoproduction~\cite{E12-19-003}.

\subsection{Impact of SRCs on other fields of physics}

While 2N-SRCs are most easily studied in high-energy scattering measurements, their impact on nuclear structure can be seen in neutron stars~\cite{Higinbotham:2009zz, Frankfurt:2008zv}, $\nu$-A scattering~\cite{Kulagin:2007ju, Niewczas:2015iea, VanCuyck:2016fab} for nuclear structure studies and $\nu$ oscillation, and understanding the nuclear quark distributions (EMC effect) through understanding the high-momentum part of spectral function~\cite{Kulagin:2004ie}. 

The impact of 2N-SRCs in nuclei can modify e-A, $\nu$-A, and A-A interactions in a way that is not captured in mean-field calculations of the nuclear structure. While \textit{ab initio} calculations of light nuclei and infinite nuclear matter can include these contributions, including the observed np-dominance, this is more challenging for medium- and heavy-mass nuclei. In addition, there are many cases where mean-field calculations that do not include SRC contributions in the scattering observables have been used, due to lack of appropriate \textit{ab initio} calculations in the past or where the mean-field structure was assumed to be sufficient.

Our present understanding of SRCs provides guidance on the overall strength and isospin structure of their contributions. In many cases, this is sufficient to allow detailed calculations or at least realistic modeling of the contribution from SRCs in other reactions. Where \textit{ab initio} calculations are not available, or where their input is not directly used in scattering calculations, momentum distributions or spectral functions can be constructed based on a combination of mean-field and SRC contributions~\cite{CiofidegliAtti:1995qe, CiofidegliAtti:2017tnm}, although there has been progress recently on extracting structure functions directly from \textit{ab initio} calculations~\cite{Wiringa:2013ala, Carlson:2014vla, Sobczyk:2021dwm}. The Generalized Contact Formalism (GCF) discussed in Sec.~\ref{sec:GCF} provides a relatively simple and flexible way of implementing SRCs in structure and scattering calculations.

\subsection{Structure of the nucleons within SRCs}

The scale of nuclear binding (tens of MeV) is so much smaller than the GeV energy scales associated with deep-inelastic scattering measurements which probe the structure function and quark distributions of nucleons and nuclei. Because of this scale separation, it was initially assumed that the quark distributions in a nucleus would be the sum of its constituent proton and neutron quark distributions, with percent level effects associated with binding and Fermi momentum.  The EMC collaboration first observed~\cite{EuropeanMuon:1983wih} that the structure function from iron different significantly from that of deuteron, taken as an approximation for the free proton and neutron distributions. This demonstrated that the quark structure of the nucleus has significant deviations from the proton and neutron distributions, and was dubbed the EMC effect. Experiments at SLAC~\cite{Gomez:1993ri} expanded these measurements to additional nuclei showing that this was a universal feature of finite nuclei, with the deviation of the nucleus from a simple sum-of-nucleons picture grew with nuclear density~\cite{Sick:1992pw}. 

While many explanations have been proposed to explain the EMC effect, as summarized in Refs.~\cite{Geesaman:1995yd, Malace:2014uea}, we do not yet have a clear understanding of the underlying physics.  Several works have examined in more detail the impact of binding and Fermi motion of the nucleons, non-nucleonic contributions to the nuclear quark distributions (e.g. those carried by virtual pions in the nucleus), and modification of the nucleon structure when bound in a nucleus. While there is a growing consensus that contributions beyond binding and Fermi motion are required~\cite{Miller:2001tg,Smith:2002ci}, additional measurements are required to better constrain attempts to explain nuclear quark distributions. 

Several experiments have been proposed to try and elucidate the origin of the EMC effect~\cite{Geesaman:1995yd, Malace:2014uea, Cloet:2019mql}. In the following section we will discuss recent ideas that were motivated by Jefferson Lab studies of both SRCs and the EMC effect, as well as the experiments that motivated these new ideas.

\subsubsection{EMC-SRC correlation}

As noted in Sec.~\ref{sec:adep}, measurements of SRCs in light nuclei~\cite{Fomin:2011ng} showed that the contribution of SRCs in light nuclei deviated significantly from the simple scaling models used to describe data in heavier nuclei~\cite{Sick:1992pw}. Models where the SRC contribution scales with average nuclear density failed to describe the anomalously large value of $a_2$ in $^9$Be, whose average density is very low, while models that scale with the target mass could not explain the large difference between $a_2$ in $^3$He and $^4$He~\cite{Arrington:2012ax}.  

The data on SRCs matched the unexpected behavior of the EMC effect~\cite{Seely:2009gt, Arrington:2021vuu} showed a non-trivial correlation between the EMC effect and presence of SRCs. Prior to this, it was generally assumed that both effects scale with the average nuclear density~\cite{Sick:1992pw,Geesaman:1995yd} , while the unexpected behavior in light nuclei showed that both the EMC effect and the presence of SRCs are sensitive to the detailed nuclear structure, suggesting that both effects are driven by the same underlying physics or that the EMC effect is driven by the presence of SRCs in nuclei. This has been referred to as the EMC-SRC correlation, and while a great deal of effort has gone into examining the correlation, it is as yet unclear what underlying physics drives the connection between these two observables.

This idea of a connection between the EMC effect and SRCs was first raised in Ref.~\cite{Higinbotham:2010ye}, which quantified the correlation between the quantities and interpreted as a consequence of the density dependence of both effects. Shortly after, Ref.~\cite{Weinstein:2010rt} examined this correlation and speculated that the EMC effect was associated with the high virtuality of nucleons in SRCs. At the time, only $^3$He, $^4$He, $^{12}$C, and $^{56}$Fe had measurements of both the EMC effect and SRC contributions, and it was already known that both effects scaled approximately with nuclear density or A$^{-1/3}$~\cite{Sick:1992pw, Gomez:1993ri}. It wasn't until the SRC data on $^9$Be~\cite{Fomin:2011ng} that there was a common deviation from the simple dependence on density for both the EMC effect and for SRCs.

Initially, the $^9$Be EMC effect was initially explained~\cite{Seely:2009gt} by looking at the local density, as seen by the struck nucleon, rather than the average density of the nucleus. This, combined with the $\alpha$-cluster structure of $^9$Be, yielded a large difference between the predication based on average density or local density (which can also be interpreted as quantifying the amount of overlap in nucleons). Ref.~\cite{Weinstein:2010rt} described the EMC-SRC correlation as a consequence of the high-momentum (large-virtuality) nucleons in SRCs. Because the SRCs are generated by nucleons interacting at short distances, there is a connection between the contribution of nucleons at short distances and the number of high-momentum nucleons, making it difficult to determine whether the EMC effect is driven by short-distance configurations, as in the local density (LD) picture, or high-momentum SRCs in the high-virtuality (HV) picture. In both cases, the cluster structure of $^9$Be yields short-distance configurations which would explain the larger-than-expected contribution of SRCs. In the HV approach, the enhanced SRC contribution causes the increased EMC effect, while in the LD picture, the short-distance, high-density clusters yield an enhanced EMC effect as well as an enhanced SRC contribution. Both pictures can explain the correlation, but may have quantitative differences, e.g. because np-dominance of SRCs would imply that only np pairs would generate the EMC effect in the HV picture, while all short-distance NN pairs can contribute in the LD model. 

Ref.~\cite{Hen:2012fm} reexamined the correlation including the data of~\cite{Fomin:2011ng} and found that the observation of the EMC-SRC correlation was largely independent of what corrections were applied to the data, e.g. to account for potential isospin structure of the SRCs. The ``robustness'' of the correlation is a consequence of the limited precision of the measurements, in particular for the EMC effect, making it difficult to use the correlation to test different hypotheses of the EMC-SRC correlation with the data available at the time. Even so, a quantitative analysis comparing simple models of the correlation for the LD and HV approaches showed a slightly better correlation in the LD picture, but the difference was only at the two standard deviation level~\cite{Arrington:2012ax}.

\subsubsection{Potential implication for flavor dependence of the EMC effect}~\label{sec:EMCflavor}

The observed EMC-SRC correlation, combined with the observation of np-dominance in SRCs suggests the possibility that the EMC effect might have an isospin dependence associated with the excess of high-momentum protons in neutron-rich nuclei. In this case, the fraction of high-momentum protons (neutrons) scales as N/A (Z/A), such that there are an equal number of high-momentum protons and neutrons. This suggests that the EMC effect in protons (neutrons) would increase (decrease) with neutron excess. This will introduce a flavor dependence of the EMC effect, with enhanced modification of the u-quark distributions in neutron rich nuclei.  Note that that a flavor-dependent EMC effect is also a natural consequence of looking individually at the proton and neutron distributions, whether you assume that the EMC effect is driven by high virtuality or local density or even different overlap between protons and neutrons, e.g. in a neutron-rich nucleus with a neutron skin~\cite{Arrington:2015wja}.

Such a flavor dependence is difficult to observe in inclusive EMC effect measurements, as the neutron excess tends to increase with the mass of the nucleus, and the existing data make it very difficult to separate an A dependence from an N/Z dependence in heavy nuclei. However, an extension of the analysis mentioned above~\cite{Arrington:2012ax} was performed, assuming that only np-SRCs contribute to the EMC effect, thus providing a specific flavor dependence for the EMC effect and allowing for the extraction of a `universal' EMC effect for the np-SRCs~\cite{CLAS:2019vsb}. They extracted their proposed universal modification function, under the assumptions noted above, and found that it was largely independent of A, concluding that was in fact a universal function with with a flavor-dependent EMC effect driven by the isospin structure of np-SRC dominance.

\begin{figure}[htbp]
\centering

\parbox{6.3cm}{
\includegraphics[width=6.3cm,height=4.8cm]{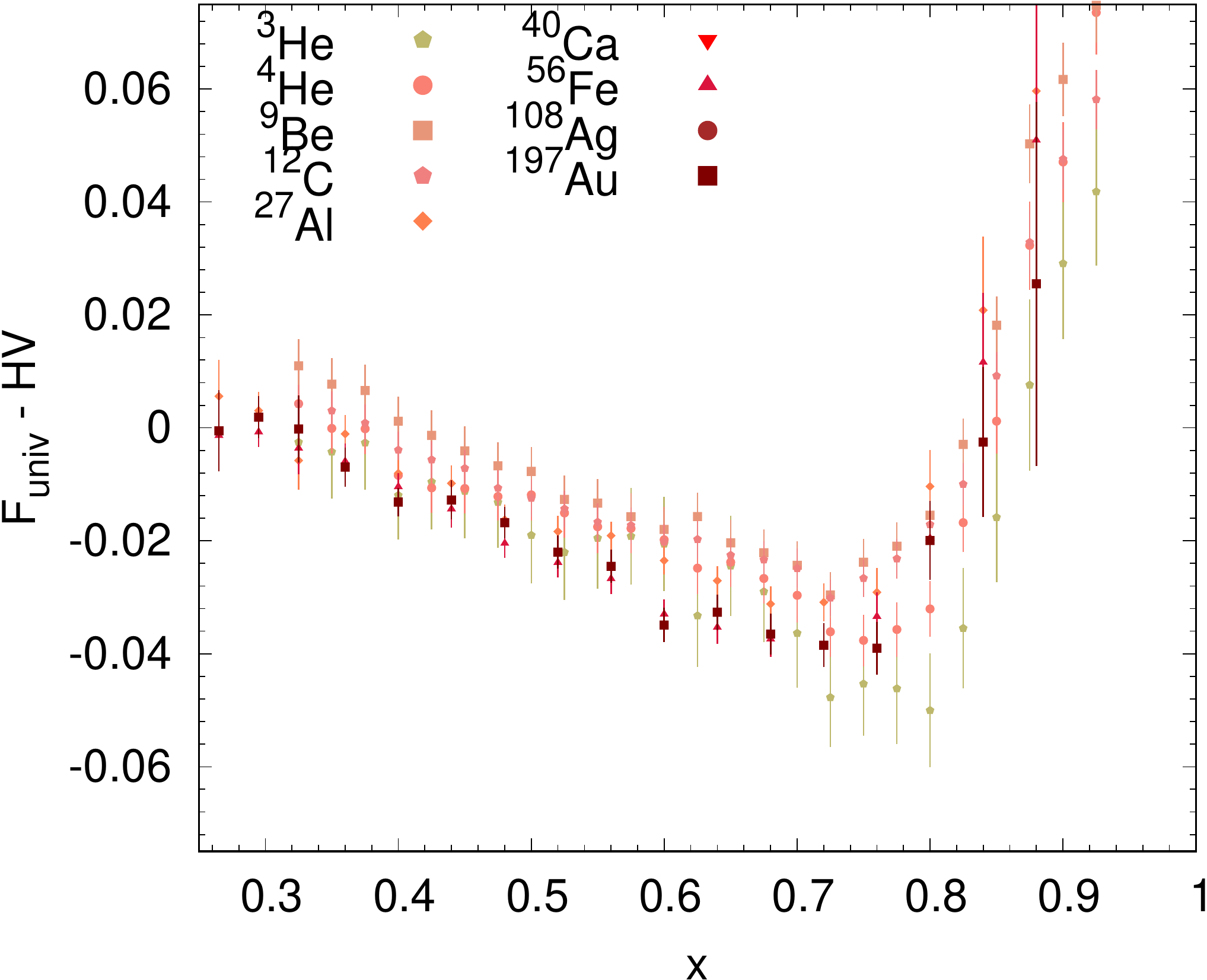}}
\parbox{6.3cm}{
\includegraphics[width=6.3cm,height=4.8cm]{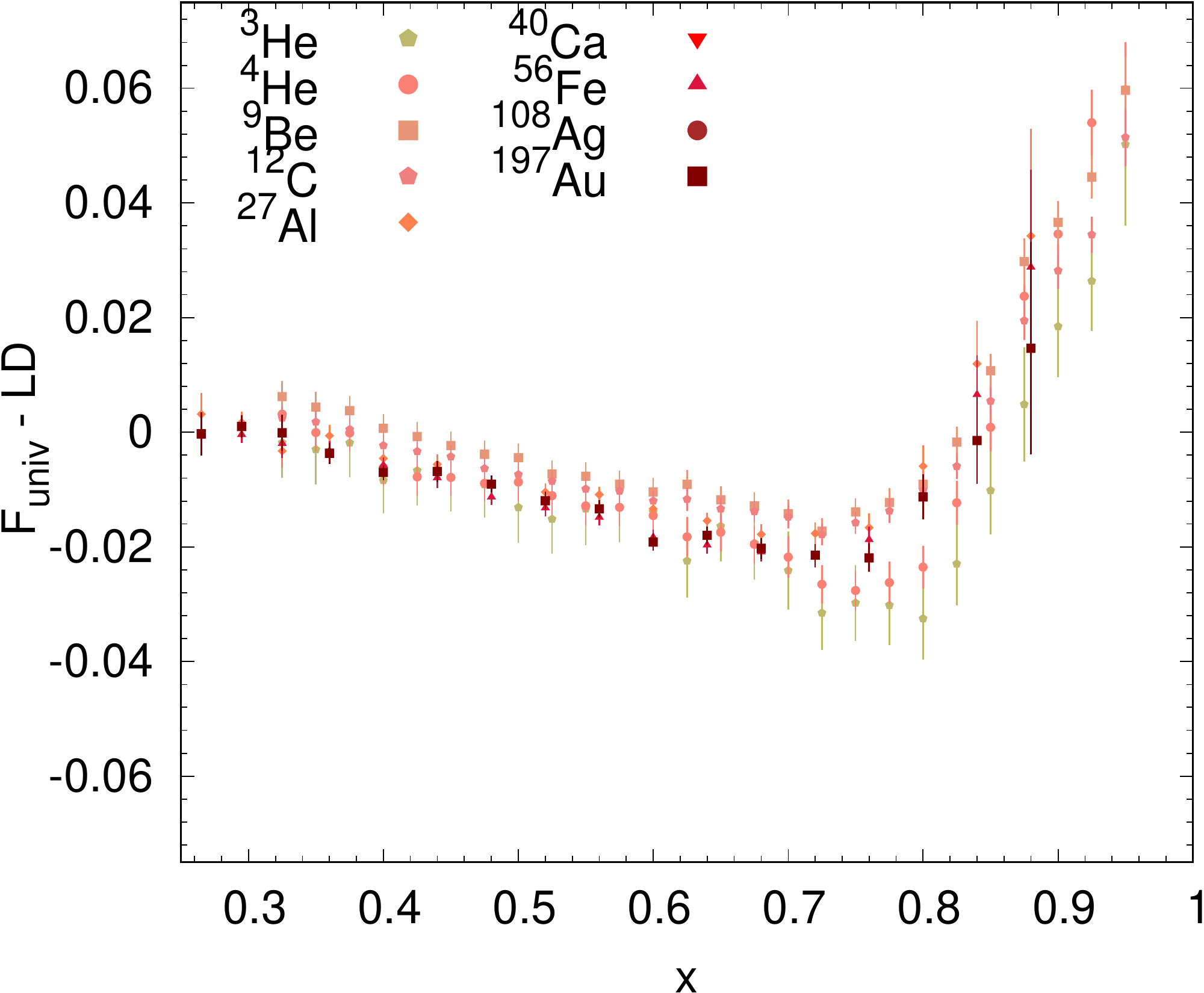}} \\
\parbox{6.3cm}{
\includegraphics[width=6.3cm,height=4.8cm]{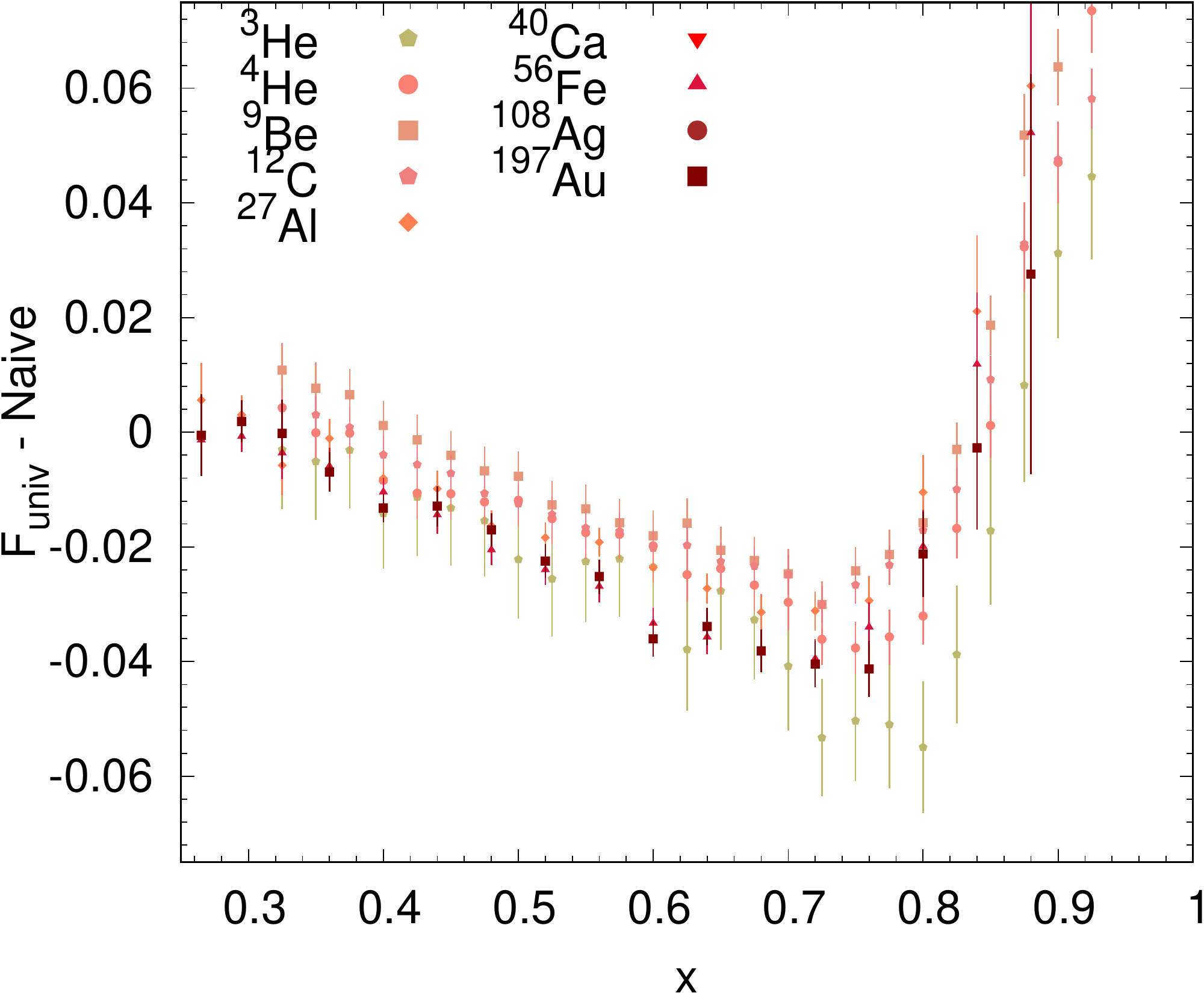}}
\parbox{6.3cm}{
\includegraphics[width=6.2cm,height=4.8cm]{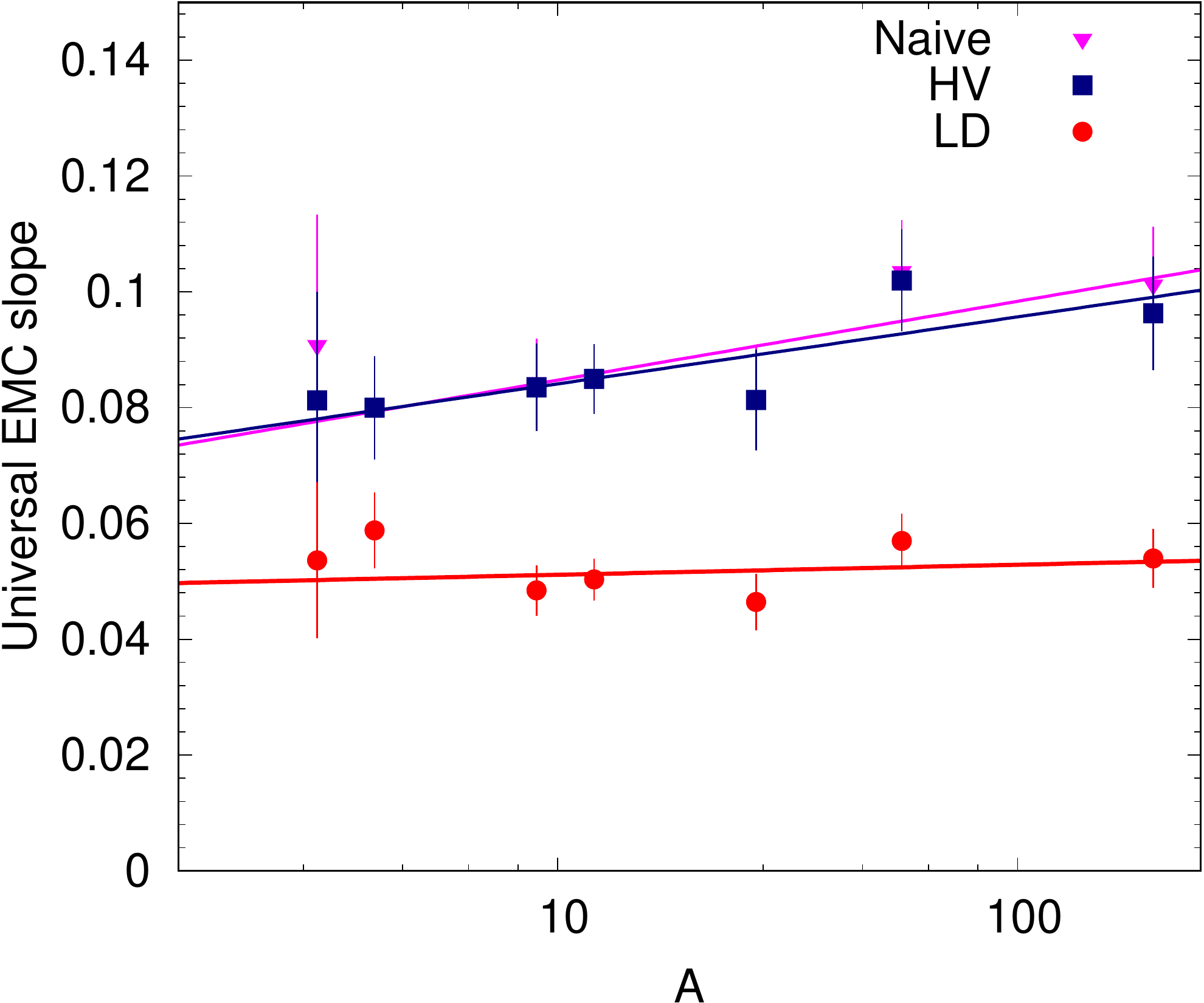}}
\caption{
Three versions of universal deuterium EMC effect. Top panels are the High Virtuality (HV) and Local Density (LD) functions of Ref.~\cite{Arrington:2019wky}, and the bottom left panel is the Naive universal function with no explicit isospin dependence (EMC effect for each nucleus scaled down by it's measured value of $a_2$). The bottom right panel compares the slope from each of the universal functions for the data from Refs.~\cite{Gomez:1993ri, Seely:2009gt, Arrington:2021vuu}.}
\label{fig:EMC-universal}
\end{figure}

Shortly thereafter, another analysis compared the universal function based on np-SRC dominance (``HV'') with an equivalent function based on the local density model (``LD''), where all NN pairs at short distance contribute (analogous to the comparison of the EMC-SRC correlation assuming HV or LD in Ref.~\cite{Arrington:2012ax}). The results are shown as the top left (HV) and top right (LD) plots of Fig.~\ref{fig:EMC-universal}. While the exact form of the isoscalar corrections applied to the data differed slightly from Ref~\cite{CLAS:2019vsb}, the universal function based on the HV assumption (np-dominance) was nearly identical to the result of~\cite{CLAS:2019vsb}. The LD-based universal function, assuming a flavor-independent EMC effect, showed a more consistent EMC slope as a function of A. Examining the A dependence of the universal functions (bottom right panel of Fig.~\ref{fig:EMC-universal}, the LD model was more consistent with the hypothesis of A-independence, but only at the two standard deviation level.

The bottom left plot of Fig.~\ref{fig:EMC-universal} shows the slopes from a third version of a universal modification function based simply on the observation of the EMC-SRC correlation. In this case the universal function is obtained by scaling down the nuclear EMC effect (deviation of the structure function ratio from unity) by the factor $a_2$, with no assumptions about the underlying cause or flavor dependence dependence. For isoscalar nuclei, this is equivalent to the HV universal function from~\cite{CLAS:2019vsb}, while for the non-isoscalar nuclei, it has an extremely small impact on the slope of the universal EMC function, as seen in the bottom right panel of Fig.~\ref{fig:EMC-universal}. As such, the data have very little direct sensitivity to the explicit assumption of flavor dependence in the HV-based universal function compared to the naive model of a flavor-independent EMC effect that scales with $a_2$. The difference between the HV and LD functions does not come directly from the explicit flavor-dependent effect, instead it comes from the fact that the LD model takes $a_2$ as a relative measure of the contribution of np-SRCs, and then scales from the number of np pairs (NZ) in the nucleus to the total number of NN pairs (A(A-1)/2). This scaling factor, A(A-1)/(2NZ), yields an A dependence even for isoscalar nuclei, and gives a reduced A dependence in the universal EMC slope. However, this makes the implicit assumption that all NN pairs contribute equally at short distances, while the short-distance distributions for pp, np, and nn pairs can differ. Thus, these comparisons are useful in examining the impact of certain assumptions on the extracted size of the EMC effect in the deuteron, but the data have limited sensitivity even in the extreme/simplified cases assumed in such analysis. Future data~\cite{E12-06-105,E12-10-008} will improve the situation by measuring the contribution of SRCs and the EMC effect for a range of light nuclei and for medium-to-heavy nuclei covering a range in N/Z, providing a data set with a greater ability to separate the A dependence and the N/Z dependence of the EMC effect.

\subsubsection{Impact on other physics}~\label{sec:EMCimpact}

The question of whether the EMC effect is flavor dependent and, if so, how large the effect is, is an important question that relates to both the origin of the EMC effect and the question of obtaining reliable quark distributions for both non-isoscalar nuclei and the neutron. The bottom right plot in Fig.~\ref{fig:EMC-universal} shows a smaller universal EMC effect for the deuteron, by almost a factor of two based on the average of all nuclei, and a factor of 1.5 if the behavior is extrapolated to the deuteron using the fits shown. This difference is significant in the extraction of the neutron structure function~\cite{Arrington:2008zh,Arrington:2011qt} and quark distributions~\cite{Accardi:2009br, Accardi:2011fa} from comparisons of proton and deuteron structure functions. In addition, it was pointed out in~\cite{Arrington:2012ax} that a flavor dependence of the EMC effect in non-isoscalar nuclei would impact the extraction of the neutron structure function from comparisons of $^3$H and $^3$He~\cite{Abrams:2021xum}, and initial examinations of the impact of such a flavor dependence using the data from~\cite{Abrams:2021xum} have already appeared~\cite{Cocuzza:2021rfn,Segarra:2021exb}. 

In addition, high-energy scattering or collider experiments (e-A, $\nu$-A, or A-A) with heavy neutron-rich nuclei. A reliable analysis of such data will require an understanding of the modification of up and down quarks in non-isoscalar nuclei.  In addition, polarized $^3$He is often used as an effective polarized neutron target, as the spins of the protons cancel almost completely. However, if used for DIS studies of the polarized pdfs of semi-inclusive DIS, a flavor dependent EMC effect will yield an additional correction currently neglected in such analyses.  Beyond this, the fact that the neutron polarization is largely associated with lower-momentum nucleons~\cite{Wiringa:2013ala} means that the flavor dependent EMC effect may differ from the unpolarized EMC effect both in their $x$ dependence and in their sensitivity to the part of the $^3$He wave function sampled in the experiment. 

Thus, the question of whether only np-SRCs or all short-distance pairs contribute to the EMC effect has a significant impact on nuclear pdfs, the predicted EMC effect for the deuteron, and spin structure studies on $^3$He. Note that the question of flavor-dependent vs flavor-independent effects is not the same as the question of local density vs high virtuality. The LD hypothesis does not require that the EMC effect be flavor independent (or flavor dependent); this is just the assumption typically made for comparison as the goal is to examine how sensitive the data is to different models of the flavor dependence. 
In addition, other ideas have been proposed to make direct measurements of the flavor dependence of the EMC effect in pion-induced Drell-Yan~\cite{Dutta:2010pg}, SIDIS~\cite{PR12-09-004,E12-21-004} and Parity-violating electron scattering~\cite{Cloet:2012td,PR12-21-002}. A review of ideas and future plans to more fully understand the EMC effect can be found in~\cite{Cloet:2019mql}.

\section{Three-nucleon short-range correlations}\label{sec:3N-SRC}

\begin{figure}[htpb] 
\parbox{6.3cm}{
\includegraphics[width=6.2cm]{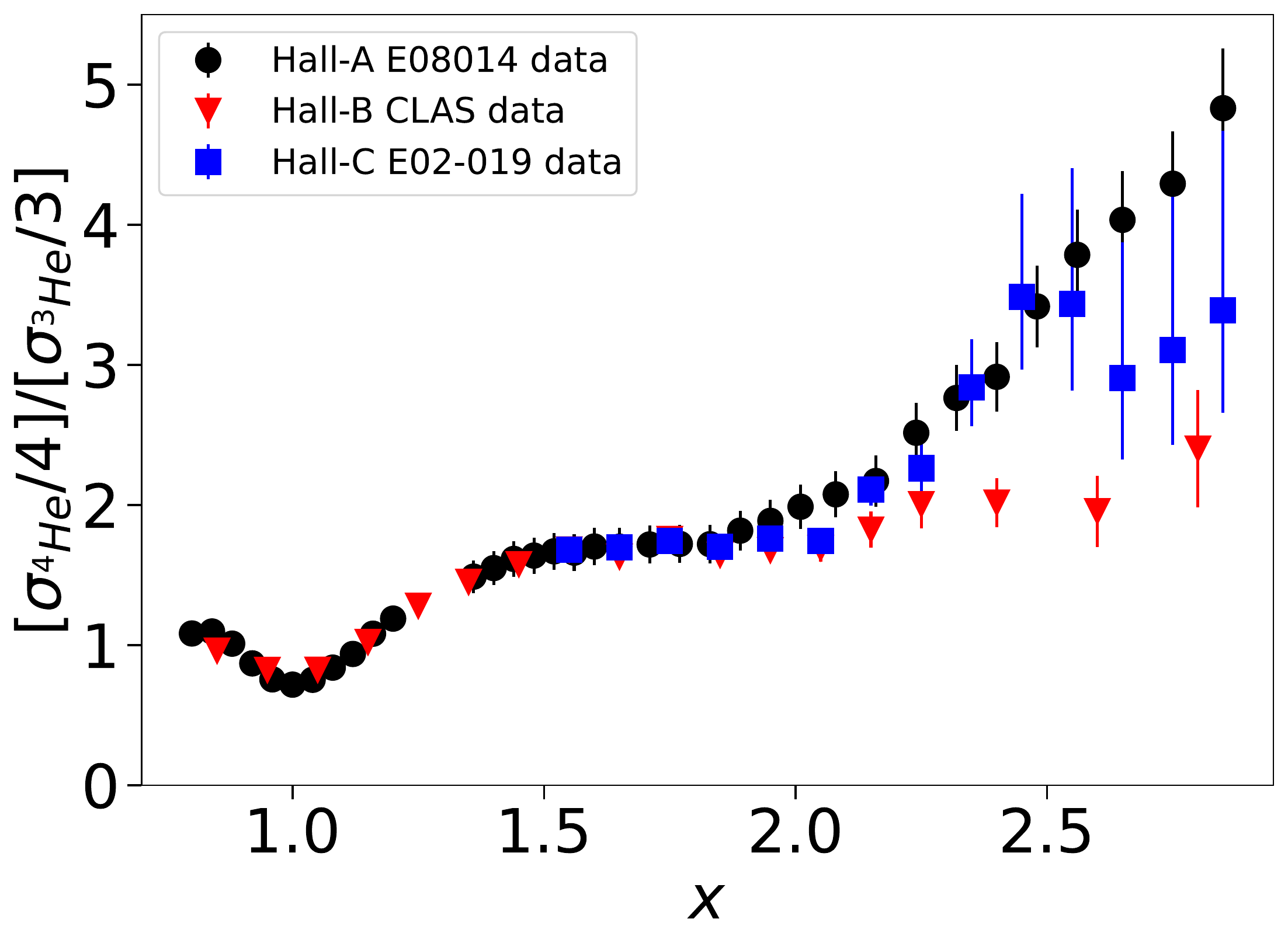}}
\parbox{6.3cm}{
\includegraphics[width=6.2cm, trim={0mm 0mm 0mm 0mm}, clip]{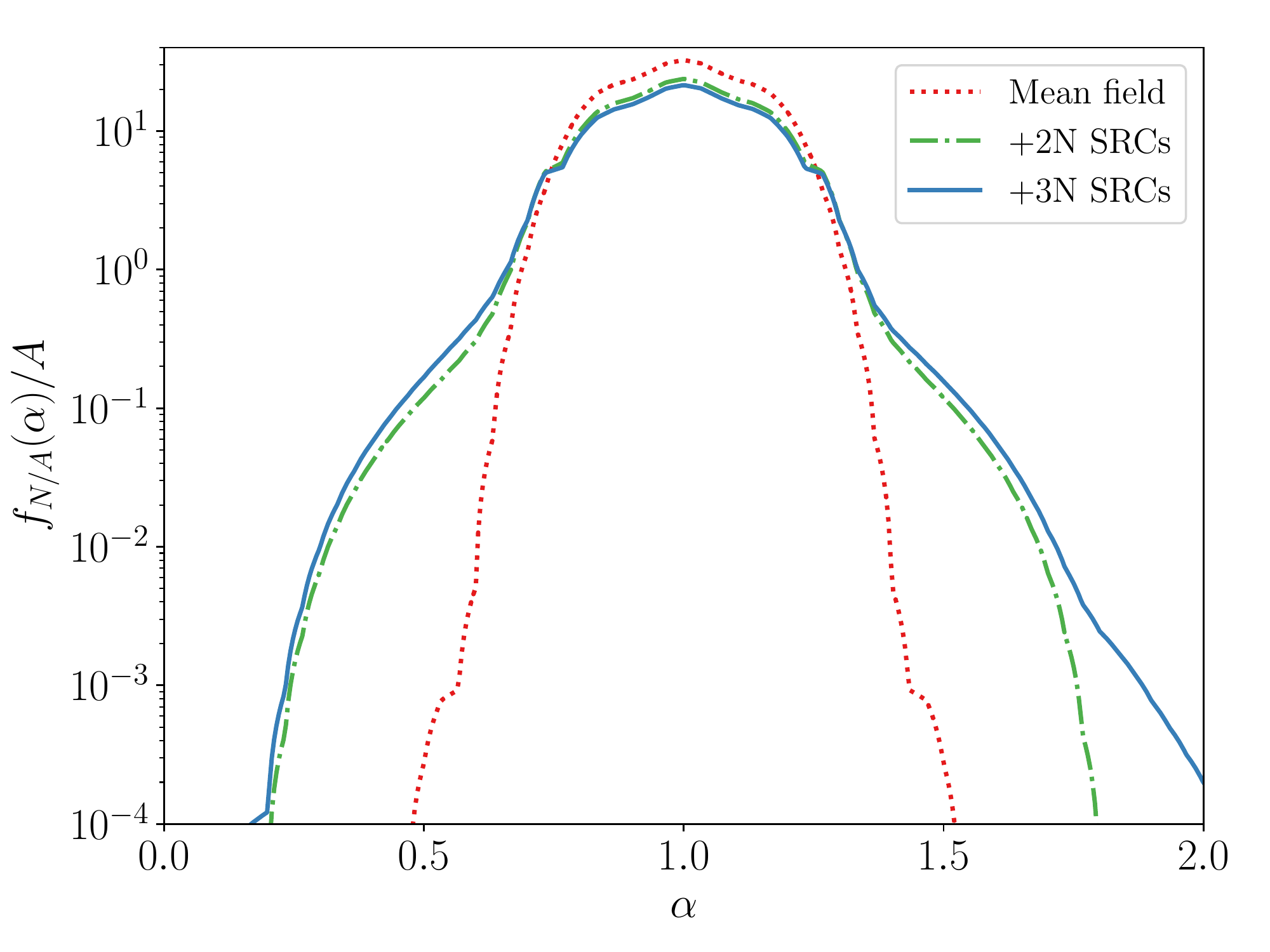}}
\caption{(Left) $^4$He/$^3$He per-nucleon cross section ratio vs $x$; figure adapted from Ref.~\cite{Ye:3Nsrc}. (Right) Calculation of the mean field, 2N-SRC, and 3N-SRC contributions from Ref.~\cite{Freese:2014zda}.}
\label{fig:3N_prev}
\end{figure}

So far, we have focused on 2N SRCs, where only two nucleons interact to produce large relative but small total momenta. Short-range configurations with more than two nucleons are not prohibited, but are expected to occur with decreasing probability.  Following searches for 2N SRCs, attempts were also made to observe a second plateau in nuclear cross section ratios relative to $^3$He. The first observation claim was published by the CLAS collaboration~\cite{CLAS:2005ola}, showing a second plateau starting at $x\approx 2.25 $, shown in Fig.~\ref{fig:3N_prev}.  This was surprising, as the $Q^2$ value was low (average of 1.7~GeV$^2$) and the onset of scaling was low in $x$. Other experiments were carried out~\cite{Fomin:2011ng, Ye:3Nsrc} with high resolution spectrometers, with the aim of looking at the $Q^2$ dependence of the second plateau as well as looking at its nuclear dependence. The data from the Hall C E02-019 experiment~\cite{Fomin:2011ng} are inconclusive, as the data are consistent with a plateau but the uncertainties in the region of interest are large.  However, the ratio significantly higher than what was observed by CLAS, and the would-be plateau starts later in $x$ (which aligns better with expectations), contradicting the interpretation of the CLAS data in terms of isolating 3N-SRCs.  A follow-up Hall A experiment~\cite{Ye:3Nsrc} performed a scan over $Q^2$ including the CLAS kinematics, but failed to observe a 3N plateau, as can be seen in Fig.~\ref{fig:3N_prev}. The plateau initially observed in Ref.~\cite{CLAS:2005ola} was later shown to be an effect of bin migration, where all the $x>2.2$ bins came from a single $E'$ bin~\cite{Higinbotham:2014xna}.  This means that we still have no definitive experimental observations of 3N SRCs and the question of whether there is a kinematic region where they dominate the scattering is still open.

\begin{figure}[htpb] 
\includegraphics[width=0.85\textwidth,height=4.5cm]{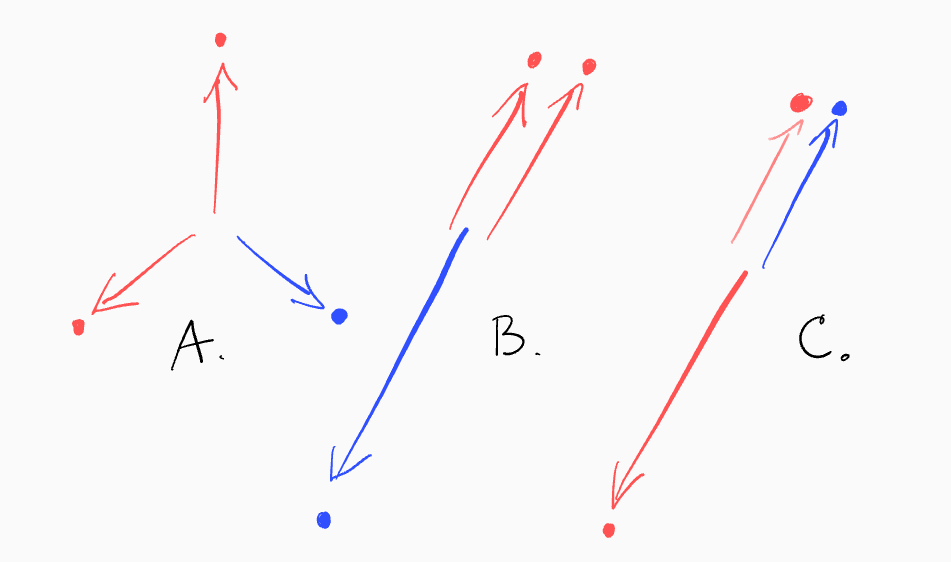}
\caption{Different potential momentum sharing and isospin configurations for symmetric (A) and asymmetric (B, C) 3N-SRC configurations.}
\label{fig:3Npic}
\end{figure}

While the onset of the 2N SRC plateau has well defined kinematics, the picture is less clear for 3N-SRCs due to multiple possible configurations, as shown in Fig.~\ref{fig:3Npic}.  Three nucleons can share three units of momentum equally (symmetric configuration), which would result in nucleons with high momenta comparable to those in a 2N configuration. Additionally, a high-asymmetric configuration is also possible. While we can calculate $p_{min}$ as a function of $x$ and $Q^2$ for heavy nuclei, this value is dependent on the momentum distribution of the spectators in the 3N-SRC, so a more realistic estimate of relevant minimum momenta for 3N-SRCs is dependent for the symmetric and asymmetric momentum configurations.

Another difficulty arises because the contribution from 2N-SRCs falls off more slowly than the mean field contributions, and so predicting what nucleon momentum is required for the 3N-SRCs to dominate (and result in a second plateau) depends on the details of size and nature of the 3N-SRC contributions. On top of this, the structure of the 3N-SRC is more complicated and predictions for its contribution are highly model dependent.  Results of one such calculation are shown on the RHS of Fig.~\ref{fig:3N_prev}, assuming the star-configuration for a 3N-SRC~\cite{Freese:2014zda}. In this model, the 3N-SRC contribution begins to have a significant contribution at $\alpha \approx 1.5$, but does not dominate until $\alpha \approx 1.8$. While the transition from mean field to 2N-SRC dominance is clear, due to the rapid and well understood falloff of the mean field momentum distribution, the transition to 3N-SRCs depends strongly on both the 2N-SRC and 3N-SRC structure assumed in the calculations.  Based on this calculation it was argued that 3N-SRCs might be visible somewhere above $\alpha=1.6$~\cite{Sargsian:2019joj}, corresponding to $x = 2.3$ for the E02-019 data~\cite{Fomin:2011ng}, with all other data sets stopping before $\alpha=1.6$. While the ratio is consistent with a plateau for $x>2.3$, the statistics are limited and data exist for only one $Q^2$, meaning that this data set is not sufficient to identify universal 3-body behavior in $^3$He and $^4$He. Ref.~\cite{Sargsian:2019joj} also predicts that $a_3(A)$, the A/$^3$He ratio in the 3N-SRC region, should go like $a_2^2$ (neglecting isospin differences), and they find that the E02-019 data are consistent with this prediction. Thus, there are hints that $\alpha=1.6$ might be sufficient to study 3N-SRCs, but new data at higher $Q^2$ will be needed to confirm the presence of 3N-SRCs~\cite{E12-06-105,LOI12-21-001}.  

The 3N-SRC landscape is further clouded by the possibility of multiple isospin configurations. If we expect some analog with the 2N SRC, where there's a region of np dominance, there should be ppn and pnn dominance among 3N-SRC configurations. Even if SRCs are dominated by ppn and pnn contributions (as opposed to 3 of a kind, which could be less probable, as pp and nn are in 2N-SRCs), scattering from isoscalar nuclei should have identical contributions from these configurations (plus possible configurations from ppp- or nnn-SRCs). Previous and upcoming measurements have used the $^3$He nucleus for the measurement in the denominator, as it is the smallest stable nucleus with three nucleons.  However, in $^3$He, only the ppn 3N-SRC configuration is possible, meaning the $a_3$ ratio will be sensitive to differences in the number of protons and neutrons at the largest momenta in 3N-SRCs. 

In a symmetric configuration (left panel of Fig.~\ref{fig:3Npic}), scattering from the highest momentum nucleons in $^3$He will involve both protons and the neutron.  In a highly asymmetric configuration (middle and right panels of Fig.~\ref{fig:3Npic}), the high-momentum part of the distribution could be dominated by the singly-occurring neutron, the double-occurring proton, or some combination. Thus, the question of whether the highest-momentum nucleons are protons, neutrons, or a mix of both, will modify the A/$^3$He ratio. If the relative contributions of protons and neutrons varies as a function of momentum, this could also distort the $x$ dependence of the ratio. Future JLab 12 GeV measurements~\cite{E12-06-105, LOI12-21-001} should shed light on many of these questions.

\section{Key questions, future directions}

In this review, we have tried to highlight recent progress in our understanding of the short-range structure of nuclei, highlighting key discoveries and important insight from theory. Here, we summarize some of these key points and discuss the remaining physics questions that will complete our picture of short-range correlations and understand their impact on nuclear structure at both the hadronic and partonic levels. Plans to expand on previous measurements and new novel probes of SRCs aimed at addressing these remaining questions will be presented.

\subsection{Summary of our understanding of SRCs and remaining questions}

Inclusive measurements identified kinematic regions of 2N-SRC dominance in nuclei, confirming the predictions for the kinematic region where SRCs should dominate and observing the expected universal two-body behavior through scaling in $x$ and $Q^2$. Such measurements mapped out the A dependence of the contribution of SRCs, identifying $^9$Be as an outlier from the expected scaling with density and indicating that understanding SRCs requires a microscopic picture of the nuclear structure. 

They also showed an intriguing correlation with measurements of the EMC effect, raising questions about the flavor dependence of the EMC effect. However, only the $^9$Be data indicate that this correlation goes beyond simple scaling of both effects with nuclear density, and the direct sensitivity of both SRC and EMC measurements to isospin structure is limited, due to the limited number of measurements on nuclei with neutron excess, and the fact that the neutron excess tends to grow with mass in heavier nuclei, making it difficult to separate mass- and isospin-dependent effects. 

Triple-coincidence measurements have given us insight into the isospin structure of 2N-SRCs, showing an enhancement of np-SRCs over pp- and nn-SRCs at the $\sim$30:1 level in all nuclei measured.  This observation has been confirmed in a variety of measurements, with new results to come for $^3$H and $^3$He~\cite{Cruz-Torres:2019fum, tritium-src}.

While 2N-SRCs have been identified, quantified, and significant progress has been made in understanding their momentum and isospin structure, we do not yet know what if any role 3N-SRCs play in nuclear structure. While early measurements suggested that they had been observed~\cite{CLAS:2005ola}, later experiments~\cite{Fomin:2011ng, Ye:2018jth} did not agree with the initial observation, which is now believed to have been limited by the experimental momentum resolution~\cite{Higinbotham:2014xna}. Because each of these experiments had one or more significant limitations - low $Q^2$ values, poor statistics, insufficient resolution - an argument can be made that there has yet to be a meaningful test of the 3N-SRC hypothesis.

Based on the discussion above, we identify the following as key questions that require further data and insight to be fully answered.
\begin{itemize}
    \item How can we improve direct, quantitative comparisons of different measurements of 2N-SRCs.  How well can we constrain calculations by combining the results of these different experiments? 
    \item What is the contribution of 3N-SRCs, and what is their internal momentum and isospin structure?
    \item What is the origin of the EMC-SRC correlation and what does it imply for the internal structure of nucleons in SRCs and the flavor dependence of the EMC effect?
\end{itemize}

\subsection{Extensions of the current experimental program}

Several experiments are planned which will extend the type of SRC studies we have discussed already. These include experiments that will make measurements on new targets, extending the kinematic coverage to higher $x$ or $Q^2$, or providing improved statistics. We summarize these experiments below, and in the following sections discuss new and novel measurements will probe SRCs and their internal structure in completely new ways, and are discussed in the following sections. 

\subsubsection{2N-SRC and EMC effect measurements in new nuclei} Two experiments are scheduled to run in 2022 that will measure A/D ratios to extract the contribution of SRCs~\cite{E12-06-105} and the EMC effect~\cite{E12-10-008} in a wide range of nuclei. This includes several light nuclei, $^{3,4}$He, $^{6,7}$Li, $^9$Be, $^{10,11}$B, and $^{12}$C to look at the A dependence of both effects in these well understood nuclei. In addition, they will study medium-to-heavy nuclei over a range of N/Z to try and disentangle the A dependence and isospin dependence of SRCs and the EMC effect. It will also provide a better data set to examine the impact of isospin structure on the EMC-SRC correlation.

\subsubsection{3N-SRCs} Experiment~\cite{E12-10-008} will also perform the first significant search for 3N-SRCs, providing the first data with large enough $Q^2$ values and high enough precision to identify a plateau in $x$ at $x>2$, as discussed in Sec.~\ref{sec:3N-SRC}. Depending on what is observed, further measurements could be performed~\cite{LOI12-21-001} to confirm the $Q^2$ independence of the scattering and to map out the A-dependence. In addition, measurements on $^3$H and $^3$He could provide a sensitive probe of the momentum/isospin structure of 3N-SRCs, even if inclusive scattering does not allow for a clean isolation of 3N-SRC contributions~\cite{E12-21-004, LOI12-21-001}. 

\subsubsection{2N-knockout} The nuclear targets program at CLAS-12~\cite{E12-17-006A}, which began data collection in 2021, will leverage the larger acceptance, higher luminosity and data rate, and increased neutron detection capabilities of CLAS-12 to pursue higher-statistics studies of two-nucleon knockout. Combined with improved modeling of reaction effects and final-state interactions this program will refine our understanding of isospin structure, momentum structure, size, and asymmetry dependencies of SRCs, with the additional goal of detecting 3N knockout events.

A number of new programs will use different probes to test the degree to which reaction effects factorize from the nuclear structure. A 2021 experiment using the GlueX spectrometer aims to identify SRCs in 2N-knockout using photoproduction~\cite{E12-19-003}. A program of inverse kinematics measurements has begun at the Joint Institute for Nuclear Research (JINR) in Dubna, using beams of nuclei incident on proton targets. The advantage of this technique is that the residual nucleus and correlated spectator nucleons have significant momentum in the lab frame, making them easier to detect and momentum analyze. For example, the detection of an intact boron nucleus in the break-up of SRC pairs in carbon has been shown to reduce apparent final state effects~\cite{BMN:2021yag}. This technique could open the possibility of studying SRCs in unstable nuclei, for instance, at the future FAIR Facility.

In addition to these measurements, which represent natural extensions of the already-completed studies, there are new measurements that offer the possibility of providing significant new information, allowing us to better understand the microscopic structure of SRCs and the origin of the EMC effect. While there are several possibilities to make new measurements that can help identify the origin of the EMC effect~\cite{Cloet:2019mql}, we provide three examples below which are more directly connected with the connection between the EMC effect and SRCs.

\subsection{New directions: partonic structure and non-nucleonic degrees of freedom}

The existing inclusive studies have demonstrated that it is possible to isolate scattering from SRCs by going to sufficiently large $x$ values at modest-to-large $Q^2$ values. By extending to much higher $Q^2$, but maintaining sufficiently large $x$, it should be possible to stay in the SRC-dominated regime while makings DIS measurements to probe the target's quark distributions. In this way, we can study the nuclear pdfs at $x>1$ and extract the distribution of so-called `super-fast quarks' in nuclei~\cite{Frankfurt:1981mk, Sargsian:2002wc}. In this region, where SRC contributions should be large, different models make very different predictions for the super-fast quark distribution. This might also provide a way to test quark-level descriptions of SRCs, e.g. where SRCs are associated with  diquark correlations between neighboring nucleons~\cite{West2020}.

\begin{figure}[htbp]
\centering
\parbox{6.3cm}{
\includegraphics[width=6.2cm, height=4.8cm, trim=30mm 135mm 12mm 41mm, clip] {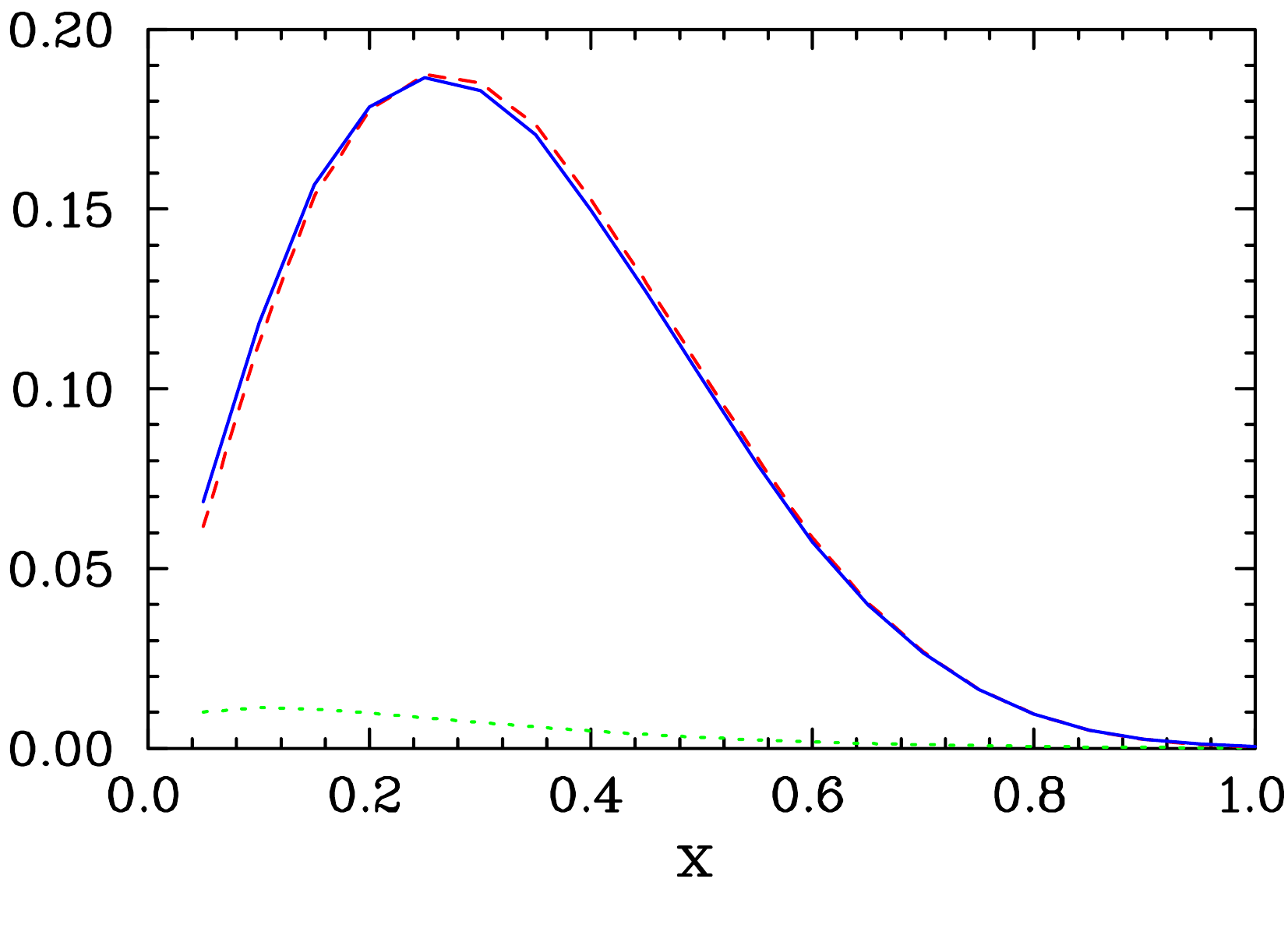}}
\parbox{6.3cm}{
\includegraphics[width=6.2cm, height=4.8cm, trim=30mm 135mm 12mm 41mm, clip] {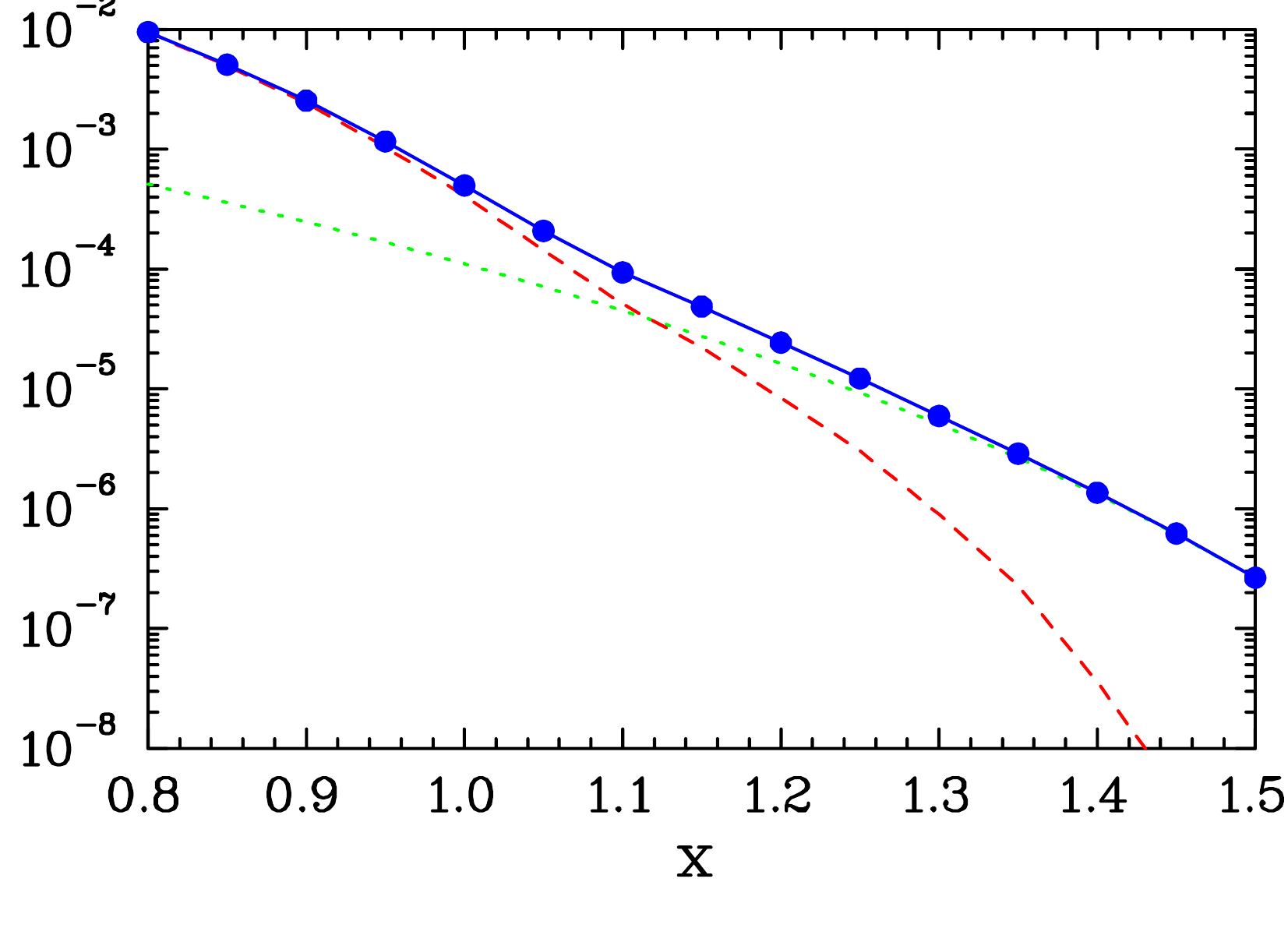}}
\caption{(Left) Plot showing the calculated valence deuteron structure function based on convolution of proton plus neutron (red dashed line) vs. the structure function (blue solid line) where the deuteron includes a 5\% component of 6q bag~\cite{Mulders:1983au} (green dotted line). (Right) Same on a logarithmic scale focused on $x>1$.}
%Published? No. Only appears in proposal (unpublished)
\label{fig:SFQ}
\end{figure}

The first such proposed measurement should take data in 2022~\cite{E12-06-105}, with $Q^2$ values from 15-20~GeV$^2$ for $x$ up to 1.4. Such a measurement should be extremely sensitive to certain types of more exotic configurations within nuclei. Early examinations of the EMC effect looked at the contribution of 6-quark bags as a model for exotic configurations for which the quark momentum distributions would be significantly modified. A small (5\%) contribution from 6q bags would have very little impact on the structure function in the DIS region, at most 1-2\%~\cite{E12-06-105}, as illustrated in the left panel of Fig.~\ref{fig:SFQ}. However, the free momentum sharing between 6 quarks carrying twice the longitudinal momentum of a single nucleon yields a large (fractional) enhancement for $x>1$, matching or even significantly exceeding the contributions that come from Fermi smearing. Thus, what is a percent-level modification for the deuteron pdf for $x<0.8$ could yield an order of magnitude enhancement over the pdfs obtained from a simple convolution of a proton and neutron, as illustrated in the right panel of Fig.~\ref{fig:SFQ}.  While this estimate was made using a 6q bag model~\cite{Bickerstaff:1984gut, Sargsian:2002wc, E12-06-105, Freese:2014zda}, any enhancement of momentum sharing between overlapping nucleons would be expected to yield a qualitatively similar effect. In models where the EMC effect is driven by highly virtual nucleons, models show a significant suppression of the structure function for off-shell as opposed to on-shell nucleons at large $x$ and light-cone momentum $\alpha$~\cite{Melnitchouk:1996vp}. DIS at $x>1$ requires $\alpha > 1$, as shown in Fig.~35 of Ref.~\cite{Hen:2016kwk}, allowing for a measurement that does not directly tag $\alpha$ for the nucleon, but which is still sensitive to the large-$x$ nucleon pdfs at $\alpha>1$, where significant suppression of the free nucleon structure function is predicted by several models~\cite{Melnitchouk:1993nk,Melnitchouk:1996vp,Hen:2016kwk}.

One critical aspect of such a measurement will be demonstrating sensitivity to the pdfs at $x>1$, where the conventional kinematics criteria for DIS scattering on a nucleon does not apply. The data taken at 6 GeV were limited to $Q^2<10$~GeV$^2$, and yet they were well reproduced by a QCD-inspired fit~\cite{Fomin:2010ei}. The 12 GeV measurement~\cite{E12-06-105} will extend these measurements to $Q^2 \approx 20$~GeV$^2$, making them more appropriate for an analysis in terms of the nuclear quark distributions. In addition, if non-nucleonic contributions yield a very large deviation from the simple proton plus nucleon convolution, the effect should be clear as long as the data is DIS-dominated, even if it is not purely DIS.

One can also make measurements that are explicitly aimed at identifying non-nucleonic degrees of freedom including hidden color configurations~\cite{Brodsky:1995rn, Brodsky:2004tq, Brodsky:2004zw} through measurements of A(e,e'N$\Delta$) or A(e,e'$\Delta\Delta$). In the presence of such non-nucleonic configurations at short-distance scales, there would be a significant enhancement of SRC-like N-$\Delta$ or $\Delta$-$\Delta$-like pairs~\cite{Ji:1985ky}. Initial estimates  suggested that such measurements would be much more sensitive at 12~GeV or even higher energies.

\subsection{New directions: Flavor-dependent EMC effect}

As noted in section~\ref{sec:EMCflavor}, the EMC-SRC connection raised the question of whether the EMC effect is flavor dependent, in particular for non-isoscalar nuclei. This will be examined by measurements of the EMC effect for non-isoscalar nuclei, e.g. $^{40}$Ca and $^{48}$Ca, but as shown in Fig.~\ref{fig:EMC-universal}, the flavor dependence associated with the neutron excess in heavy nuclei has only a modest impact on the EMC effect. While a direct measurement of sufficient sensitivity does not appear to be possible, the systematic study of JLab E12-10-008~\cite{E12-10-008} may allow for a model-dependent separation of A and isospin dependence. 

The use of SIDIS as a way of flavor tagging scattering on nuclei has been proposed as a way to examine the flavor dependence of the EMC effect~\cite{PR12-09-004,E12-21-004}. However, measurements in heavy nuclei, where the expected effects are expected to be relatively significant, have to deal with model-dependent corrections and other possible nuclear effects which may make it difficult to cleanly isolate the flavor dependence of the nuclear pdfs. These issues can be addressed by going to light nuclei, in particular in a comparison of $^3$H and $^3$He, but it is not clear that realistic models of the flavor dependence will be large enough to measure in such experiments. 

A clean and sensitive measurement can be performed using parity-violating electron scattering on non-isoscalar nuclei ~\cite{Cloet:2012td}. This involves looking for a modest change in a parity-violating asymmetry of order 100 parts-per-million, and while this is within the capabilities of modern parity-violating measurements, it requires a high luminosity and large acceptance as well as good control of false asymmetries. This appears to be within the capabilities of the SoLID detector planned for parity-violating DIS measurements at Jefferson Lab~\cite{PVDIS}, and a proposal was submitted to perform such a measurement on $^{48}$Ca~\cite{PR12-21-002} that can provide excellent sensitivity to estimates of the flavor dependence based on simple scaling models~\cite{Arrington:2015wja} or the calculations of Ref.~\cite{Cloet:2012td}.

\subsection{New directions: Tagged measurements of SRCs}

Another approach related to flavor dependence is the technique of spectator tagging, especially in deuterium, where by the detection of a recoiling spectator nucleon can identify the struck-nucleon, and provide information about the deuteron's initial configuration (i.e., short range vs. long range). The requirement of nucleon detection makes the technique sensitive to final state interactions, though theoretical estimates indicate the effects are smallest when the spectator recoils in the backwards direction~\cite{Cosyn:2017ekf,Strikman:2017koc}. This technique is used by the BoNUS experiment in CLAS, in which the detection of a low-momentum spectator proton indicates that the accompanying DIS reaction took place on a neutron. By extrapolating to stationary neutron limit, BoNUS aims to determine the structure of the free neutron. First results were published in 2011 from the 6 GeV program~\cite{CLAS:2011qvj,CLAS:2014jvt,Griffioen:2015hxa}. A subsequent 12~GeV measurement is currently under analysis~\cite{E12-06-113}.

Whereas BoNUS is focused on slow-moving spectators, two other Jefferson Lab experiments aim to detect high-momentum (300--600 MeV$/c$) spectators, as a way of determining the structure modification in SRC-pairs, thus testing models of the EMC effect. The BAND Experiment~\cite{E12-11-003A} took data in CLAS in 2019 with a backward-angle detection system added to measure backwards recoiling spectator neutrons. The Hall C LAD Experiment~\cite{E12-11-107} will detect recoiling spectator protons. Both experiments aim to determine how nucleon structure modification changes with virtuality. Their combined results may give an indication of how bound proton modification differs from that of the bound neutron. Such studies can be extended to higher energy and larger spectator momenta at the proposed Electron-Ion Collider~\cite{Jentsch:2021qdp}.

\subsection{Summary and Conclusions}

Over the past few decades, extensive sets of data related to short-range correlations have become available, mainly from the Jefferson Lab experimental program. With this data in hand, we have mapped out the contribution of SRCs over a range of nuclei, and measured their isospin and momentum structure. Overall, enough has been learned that we are in a position to incorporate this short-distance physics into nuclear structure calculations related to low energy scattering, neutrino-nucleus scattering, and neutron star structure. These new data have driven significant experimental progress in quantifying SRCs and understanding their internal structure, although more work is required to understand final-state interactions and the momentum structure of np and pp SRCs well enough make detailed, quantitative comparisons of different measurements that are sensitive to SRCs. Beyond this, the issue of 3N-SRCs remains unanswered: what is their isospin and momentum structure, and can they be isolated experimentally? These questions are important enough that studies SRCs and related high-energy nuclear structure measurements continue to be a significant part of the future Jefferson Lab experimental program~\cite{Arrington:2021alx}.

Beyond the direct measurements of SRCs in nuclei, this experimental program has also raised questions that touch on other aspects of high-energy nuclear structure. The observation of the EMC-SRC correlation raises the question of whether this represents a commonality of origin or a direct causal connection where SRCs generated modified nucleon structure because of their short-distance and/or high-momentum nature.  The possibility of a causal connection, combined with the strong np dominance of 2N-SRCs, suggests the possibility of a flavor-dependent EMC effect in non-isoscalar nuclei. This motivates further studies of the EMC effect, as well as a program of studies to probe the internal partonic structure of SRCs, either through DIS measurements at kinematics that isolate scattering from SRCs, or from tagged measurements which look at the nucleon structure (effective form factors or pdfs) as a function of nucleon virtuality.  

Answering these questions will require not only the experimental program described in this paper, but also improvements in the theory necessary to interpret these measurements.  Improved FSI corrections are needed for more quantitative studies in one- and two-nucleon knockout reactions, and the extraction of the modification of the nucleon structure through tagged measurements of effective form factors or pdfs relies on careful comparison to calculations which account for both the conventional nuclear effects and modification of the nucleon structure within SRCs.  Close collaboration has driven the simultaneous advancement of both experimental data and theoretical understanding, and will continue to open up new avenues for future study.

\textbf{Acknowledgements:} The authors wish to thank Wim Cosyn, Adam Freese, Shujie Li, Augusto Machiavelli and Jerry Miller for useful discussions and other contributions.

This work supported in part by the Department of Energy's Office of Science, Office of Nuclear Physics, under contracts DE-AC02-05CH11231, DE-SC0013615 and DE-SC0016583.

\bibliographystyle{ar-style5}
\bibliography{references}

\end{document}